# Pareidolic Illusions of Meaning: *ChatGPT, Pseudolaw and the Triumph of Form over Substance*

**Dr Joe McIntyre**

The early 20's has seen the rise of two strange, and potentially quite impactful social phenomena: pseudolaw (where users rely upon pseudolegal arguments that mimic the form and ritual of legal argumentation but fundamentally distort the content of law), and generative AI/LLMs (which generate content that uses probabilistic calculations to create outputs that look like human generated text). This article argues that the juxtaposition of the two phenomena helps to reveal that they both share two fundamental traits: both elevate form and appearance over substance and content, and users of both routinely mistake the form for the substance. In drawing upon legal theory, computer science, linguistics and cognitive psychology, the article argues that both phenomena rely upon creating illusions of meaning that users mistake for the underlying primary phenomenon. I then explore four implications of this conception of both phenomena: (1) in both cases rely on human tendencies of 'conceptual pareidolia' resulting in the erroneous perception of meaningful linguistic/legal patterns from nebulous inputs; (2) both rely upon the 'confidence heuristic', the human cognitive bias for treating confidence as a proxy for competence; (3) both phenomena succeed when the primary concern is with the form of the output and not its content; and (4) both phenomena rely heavily upon the magical thinking of users, and the desire for the promise of the approach to be real. It argues that the legal context helps to reveal a solution for the problems caused by both phenomena: it is only where users possess sufficient legal and technological literacy that it becomes possible to reveal to them the illusionary nature of the phenomena.

## Table of Contents

# PART I: INTRODUCTION

Sometimes, the best way to understand something is through juxtaposition, by contrasting it to something else. This article applies this technique to two emerging social phenomena – pseudolaw and generative AI (specifically, large language models) – that may not appear to have much in common other than becoming prominent in the last few years. Yet, when the two phenomena are studied together, their shared common attributes not only become apparent, but insight into each becomes possible that would not otherwise be apparent.

'Pseudolaw' refers to the emerging practice of people using novel, if not outright nonsense, legal arguments that mimic the form and ritual of legal argumentation but fundamentally distort the content of law.[1] Since the start of the global pandemic, these forms of argument have become increasingly common globally as disaffected citizens confront the authority of the state.[2] In contrast, Large Language Models ('LLMs') are a form of computer program which 'use reams of available text and probability calculations in order to create seemingly-human-produced writing'[3] and which have 'become increasingly sophisticated and convincing over the last several years'.[4] Other than this temporal proximity, these phenomena appear to bear no relationship to each other.

Yet the juxtaposition in this article reveals that both are a form of technology (one legal, the other computational) that share two fundamental common traits: they elevate form over substance, and they rely on users mistaking the illusory form for meaningful substance. In exploring the nature and implications of these traits, this article aims to provide insight into the core nature of both phenomena in a way that would not be possible were they analysed in isolation.

This article draws upon a range of diverse disciplines – including law, computer science, linguistics and psychology – to better understand these two emergent phenomena. Ultimately, it argues that if we are to understand these phenomena, it is necessary to understand them as complex social phenomena, involving both the underlying technology and the ways in which humans respond to it.

This article strives to be accessible and useful to readers from all the above disciplines (and beyond), because if we are to properly understand and respond to these phenomena, it will be necessary to move beyond strict disciplinary boundaries. And we should be under no doubt as to the necessity of properly responding to both phenomena, as both have the potential to profoundly disrupt entire industries and broad swathes of society.

## 1. Why Law? The Benefit of a Legal Lens to Understand Computational Technology

Of course, to those coming to this article from another discipline, the question to be addressed is: *Why Law?* It may appear a strange domain from which to address the nature of LLMs. Yet there are good reasons why law is a particularly useful lens.

---

[1] Harry Hobbs, Stephen Young and Joe McIntyre, 'The Internationalisation of Pseudolaw: The Growth of Sovereign Citizen Arguments in Australia and Aotearoa New Zealand' (2024) 47(1) *University of New South Wales Law Journal* 309 ('The Internationalisation of Pseudolaw').

[2] Ibid.

[3] Michael Townsen Hicks, James Humphries and Joe Slater, 'ChatGPT is Bullshit' (2024) 26(2) *Ethics and Information Technology* 38, 37.

[4] Ibid.

Firstly, there is a particular visibility in public discourse on the sometimes strange and memorable interplay between law and computational technology. The most famous of these was 'Lawyer Cat', the viral video of a lawyer who – in the early days of the pandemic – attempted to zoom into an online court hearing only to have his appearance thwarted by a cat 'filter' on the video stream.[5] The immortal words 'I am here live. I am not a cat',[6] have lived on in infamy. Every year Judge Roy Ferguson, the judge in the video, takes to social media to celebrate '#Lawyercat Anniversary'.[7]

Perhaps less humorously, but more newsworthy, have been the increasingly common media stories of judges[8] and lawyers[9] misusing generative AI technology, particularly Large Language Models ('LLMs') such as ChatGPT. In the last two years, these cases have gone from being something of a joke to becoming a serious concern for the administration of justice. The first of these cases to gain widespread attention occurred in late May 2023, when a lawyer in New York filed documents, citing a number of non-existent cases as authority.[10] It was subsequently revealed that the lawyer had used ChatGPT in helping to draft the brief and was 'unaware that its content could be false'.[11] A year later, Michael Cohen – Donald Trump's former personal lawyer and 'fixer' – narrowly avoided sanctions after similarly relying upon artificial intelligence-generated legal case citations.[12] The judge in that case only determined not to order sanction only after he accepted that Cohen believed ChatGPT 'to be a 'super-charged search engine' rather than a 'generative text service.'[13] Yet, lawyers around the globe continue to misuse this technology, much to the chagrin of judges.[14] Judges are now being forced to refer such lawyers to regulators for potential disciplinary action,[15] and to issue special Practice Directions to regulate the appropriate use of the technology.[16] Law is a domain where the misuse of technology is already creating profound challenges.

---

[5] 'US Lawyer Accidentally Turns on Cat Filter During Court Hearing via Zoom', *ABC News* (online, 10 February 2021) <https://www.abc.net.au/news/2021-02-10/us-lawyer-accidentally-turns-on-cat-filter-during-court-hearing/13139990>.

[6] 394th District Court of Texas – Live Stream, 'Kitten Zoom Filter Mishap' (YouTube, 10 February 2021) <https://www.youtube.com/watch?v=KxlPGPupdd8>.

[7] @judgefergusontx.bsky.social (Judge Roy Ferguson) (BlueSky, 10 February 2025) <https://bsky.app/profile/judgefergusontx.bsky.social/post/3lhqvutie7k2g>.

[8] In the Netherlands: Caroline Hill, 'Dutch Judge causes Storm by using ChatGPT for Fact Checking in Judgment', *Legal Insider* (online, 8 August 2024) <https://legaltechnology.com/2024/08/08/dutch-judge-causes-storm-by-using-chatgpt-for-fact-checking-in-judgment/>; In the US: Nate Raymond, 'US Judge Runs 'Mini-Experiment' with AI to help decide case', *Reuters* (online, 7 September 2024) <https://www.reuters.com/legal/transactional/us-judge-runs-mini-experiment-with-ai-help-decide-case-2024-09-06/>; In Chile: Luke Taylor, 'Colombian Judge says he used ChatGPT in Ruling', *The Guardian* (online, 3 February 2023) <https://www.theguardian.com/technology/2023/feb/03/colombia-judge-chatgpt-ruling>.

[9] Josh Taylor, 'Australian Lawyer Caught Using ChatGPT Filed Court Documents Referencing 'Non-existent' Cases', *The Guardian* (online, 1 February 2025) <https://www.theguardian.com/australia-news/2025/feb/01/australian-lawyer-caught-using-chatgpt-filed-court-documents-referencing-non-existent-cases>; Damien Carrick and Sophie Kesteven, 'This US lawyer used ChatGPT to Research a Legal Brief with Embarrassing Results. We Could all Learn from his Error', *ABC News* (online, 24 June 2023) <https://www.abc.net.au/news/2023-06-24/us-lawyer-uses-chatgpt-to-research-case-with-embarrassing-result/102490068>; Molly Bohannon, 'Lawyer Used ChatGPT In Court – And Cited Fake Cases. A Judge is Considering Sanctions', *Forbes* (online, 8 June 2023) <https://www.forbes.com/sites/mollybohannon/2023/06/08/lawyer-used-chatgpt-in-court-and-cited-fake-cases-a-judge-is-considering-sanctions/>; Cam Wilson, 'People are Citing Non-existent Cases and giving Rambling Statements in Court. Judges Blame AI', *Crikey* (online, 13 September 2024) <https://www.crikey.com.au/2024/09/13/court-judges-chatgpt-generative-artificial-intelligence>; *Mavundla v MEC: Department of Co-Operative Government and Traditional Affairs KwaZulu-Natal (7940/2024P)* [2025] ZAKZPHC 2.

[10] Benjamin Weisser, 'Here's What Happens When Your Lawyer Uses ChatGPT' New York Times (online, 27 May 2023) <https://www.nytimes.com/2023/05/27/nyregion/avianca-airline-lawsuit-chatgpt.html>; Damien Carrick and Sophie Kesteven, 'This US lawyer used ChatGPT to Research a Legal Brief with Embarrassing Results. We Could All Learn from His Error', *ABC News* (online, 24 June 2023) <https://www.abc.net.au/news/2023-06-24/us-lawyer-uses-chatgpt-to-research-case-with-embarrassing-result/102490068>.

[11] Kathryn Armstrong, 'ChatGPT: US Lawyer Admits using AI for Case Research', *BBC News* (online, 28 May 2023) <https://www.bbc.com/news/world-us-canada-65735769>.

[12] The Associated Press, 'Michael Cohen Says he Unwittingly Sent AI-Generated Fake Legal Cases to his Attorney', *npr* (online, 30 December 2023) <https://www.npr.org/2023/12/30/1222273745/michael-cohen-ai-fake-legal-cases>.

One of the reasons that these events attract such attention is that law is a specialised field of knowledge that is perceived as important but mysterious. Lawyers being laid low by technology inverts expectations and generates a degree of schadenfreude.

Yet more substantively, law provides an ideal context to examine the nature and limitations of LLMs. Law is a discipline that produces reams of authoritative text and which prides itself on being coherent and predictable, presenting logical and syllogistic reasoning.[17] It seems entirely consistent with this framing to think that the predictive computational models of LLMs ought to be particularly well-suited to the development of legal content. There are already a number of studies that suggest these models can produce content of sufficient quality to pass law exams.[18] Yet, ultimately, formalist articulations of legal methods are deeply misleading, and these apparent successes are deceptive. In reality, law is a deeply evaluative and discursive enterprise,[19] that is as much about Judge as Solomon as Judge as Machine.[20] These limitations of the legal discipline become significant in the way they help reveal the limitations of LLMs.

But perhaps most importantly, law provides a vital site for analysis because it is also a site where pseudolaw is presenting another phenomenon where form is elevated above substance. As I will discuss below, our legal institutions are being increasingly impacted by the rise of pseudolaw. And interestingly, evidence is beginning to suggest the interaction between pseudolaw and LLMs. Recently, anecdotal stories have begun to emerge of litigants-in-person (sometimes referred to as 'self-represented,' or 'pro se' litigants) using LLMs in court rooms to provide real-time answers to judicial questions. These submissions sound like they are legally meaningful, yet they are ultimately devoid of genuine substance. They are pseudolaw, but they are being generated by LLMs. This suggests that there may be something fundamentally common to both phenomena that is allowing them to blend seamlessly into each other.

---

[13] Ibid; Josh Russell, 'Judge Won't Sanction Michael Cohen Over AI-Generated Fake Legal Cases', *Courthouse News Service* (online, 20 March 2024) <https://www.courthousenews.com/judge-wont-sanction-michael-cohen-over-ai-generated-fake-legal-cases/>; For the judgment itself, see <https://www.courthousenews.com/wp-content/uploads/2024/03/michael-cohen-sanctions-opinion.pdf>.

[14] For a useful overview of the issue, see Brenda Tronson, 'ChatGPT Is Not A Paralegal: The Professional Implications For Lawyers In Using ChatGPT', *Australian Public Law* (online, 16 February 2024) <https://www.auspublaw.org/blog/2024/2/chatgpt-is-not-a-paralegal-the-professional-implications-for-lawyers-in-using-chatgpt>.

[15] Naomi Neilson, 'ChatGPT Blunder Sees Lawyer Referred to Regulator', *Lawyers Weekly* (online, 4 February 2025) <https://www.lawyersweekly.com.au/biglaw/41421-experienced-lawyer-referred-to-regulator-over-chatgpt-blunder>.

[16] Claudia Williams, 'Want to Use ChatGPT to Help Prepare for Court? This is What Lawyers Say you Should and Shouldn't Do', *ABC News* (online, 23 May 2024) <https://www.abc.net.au/news/2024-05-23/generative-ai-chatbots-responsible-use-in-court-guildelines/103863968>.

[17] I have previously described this as the 'methodology myth': Joe McIntyre, 'The Six Myths of Judicial Independence' (2025) 52(2) *University of Western Australia Law Review* 157, 162.

[18] See for example Jonathan Choi, Kristin Hickman, Amy Monahan and Daniel Schwarcz, 'ChatGPT Goes to Law School,' (2022) 71(3) *Journal of Legal Education* 387; Mohanad Halaweh 'ChatGPT in Education: Strategies for Responsible Implementation, Contemporary Educational Technology' (2023) 15(2) *Contemporary Educational Technology,* ep421; Miriam Sullivan, Andrew Kelly, Paul McLaughlan, 'ChatGPT in Higher Education: Considerations for Academic Integrity and Student Learning' (2023) 6(1) *Journal of Applied Learning and Teaching* 31–40; Damian Curran, Inbar Levy, Meladel Mistica & Eduard Hovy, 'Persuasive Legal Writing Using Large Language Models' (2024) 34(1) *Legal Education Review* 184; Amanda Head and Sonya Willis, 'Assessing Law Students in a GenAI World to Create Knowledgeable Future Lawyers' (2024) 31(3) *International Journal of the Legal Profession* 293; Francine Ryan and Liz Hardie, 'ChatGPT, I Have a Legal Question? The Impact of Gen AI Tools On Law Clinics and Access to Justice' (2024) 31(1) *International Journal of Clinical Legal Education* 166.

[19] Joe McIntyre, *The Judicial Function: Fundamental Principles of Contemporary Judging* (Springer, 2019) 89-91.

[20] Ibid, 87-89.

This article aims to expose the fact that there is indeed such commonality. At their core, both succeed because they create illusions of meaning.

## 2. Structure of the Article

This article aims to provide an insight into the core nature of two emergent social phenomenon: (1) The Rise of Pseudolaw; and (2) The Rise of Generative AI & LLMS. It argues that both pseudolaw and LLMs operate through the ascendency of form over substance and depend upon users mistaking the illusory form for underlying substantive meaning.

In doing so, I argue that both phenomena share a common trait: users mistake the form for the thing itself, finding meaning and patterns where none exist. For pseudolaw, this means adherents mistake pseudolegal arguments that share some sources and forms with legal reasoning with law itself.[21] Adherents perform legalistic rituals and use legalistic-sounding language, and believe that in doing so they are doing law. For LLMs that stitch together linguistic forms without any reference to meaning, humans tend to mistakenly assume that, given the familiar form, the output has meaning. The output has the *form* of language, so we assume it is engaging in communication - *a transfer of meaning*. But this is an error. All meaning is projected from the human onto the outputted form. LLMs do not produce meaning. LLMs produce pseudo-meaning.

In both cases, users mistake the illusion of meaning for meaning itself. This can be seen as a form of 'conceptual pareidolia'. As I explain below, pareidolia is the tendency for human perception to impose a meaningful interpretation on a nebulous stimulus[22] – a face on the moon, an animal in the clouds, etc. This tendency appears to be hardwired into the human brain.[23] Though pareidolia normally refers to visual perception, we can see a similar process happening in the context of LLMs and pseudolaw – diffuse inputs are erroneously reconstructed by human perception as conveying meaning when none is present. This can be understood as 'conceptual' or 'abstractive' pareidolia. I argue that by understanding these phenomena in this way, it becomes possible to better explain the misuse of digital technology of LLMs and the 'legal technology' of pseudolaw. In both cases, there is an illusion that is perceived to convey meaning.

In this paper, I draw upon the context of 'law' to provide insights into the nature, limits and implications of both LLMs and pseudolaw. As I outline above, 'law' provides a particularly rich context to illustrate the concept of 'meaning' in these domains, and the ways in which in both cases the phenomenon involves an output that tends to be mistaken by human users as contextually meaningful.

---

[21] Harry Hobbs, Stephen Young and Joe McIntyre, 'The Internationalisation of Pseudolaw: The Growth of Sovereign Citizen Arguments in Australia and Aotearoa New Zealand' (2024) 47(1) *University of New South Wales Law Journal* 309.

[22] See below Part IV(1)

[23] See Carl Sagan, *The Demon-Haunted World* (Ballantine Books, 1996) 45.

The nature of law – while of course endlessly debated by juris prudes[24] - can be best considered as a social norm distinguished by its 'higher degree of clarity, formalisation, and binding authority'.[25] Law is, inherently, a discursive enterprise[26] that exists in a social context. It involves particular systems of reasoning, is institutional and hierarchical, and is dependent upon particular forms, structures and processes. Critically, law relies upon a number of language-rich artifacts – statutes, caselaw, regulations, and secondary analysis – that create an extraordinarily large corpus of written sources. This can give the illusion that law possesses an 'objective' and ascertainable existence. Yet, law is inherently an entirely artificial construct derived from the human mind, dependent in all application on acts of evaluative judgment. Despite the reams of legal data, there is conceptually no correct legal outcome, external 'right' answer capable of validation.[27] The meaning of law is always contested and contestable. As Hart famously described, law (like language itself) is open textured[28] and essentially defeasible.

Yet the very complexity of law makes it essentially inaccessible to the majority of the population to such an extent that, too commonly, citizens are unaware of the level of their lack of legal literacy.[29] The general member of the public can not only not afford legal advice, but they lack the knowledge and skills to distinguish between legally legitimate and fanciful claims. Both LLMs and pseudolaw offer an apparent solution to this problem: the use of language and legal form that becomes accessible and powerful to the public. Already, stories are emerging of litigants using LLMs in courtrooms in real-time, producing claims that are essentially pseudolegal in nature. The lack of accessibility of law is leading people to turn to both phenomena to counter alienation.[30]

Yet, at its core, law is about a common set of agreed norms, about a communal pursuit of a shared set of meaningful rules. Unless one properly understands how the corpus of written legal artifacts is incorporated into the institutions of legal governance through discursive and participatory behaviours, it is impossible to make use of (or meaningful predictions about) law. Law is a site where form is essential, but ultimately substance is critical. Yet the very inaccessibility of law makes it particularly vulnerable for the laity to mistake form for substance.

The juxtaposition of LLMs and pseudolaw in this paper generally – and in the context of law in particular - is useful because the contrast reveals something important about both phenomena. Moreover, it invites us to reflect upon the nature of the underlying domain of law itself.

---

[24] The question 'what is law' has been asked by priests and poets, philosophers and kings and is arguably 'as old as philosophy itself': H.J Abraham, *The Judicial Process: An Introductory Analysis of the Courts of the United States, England and France* (Oxford University Press, 7th ed 1998) 2. For a good introduction, see Nigel Simmonds, *Law as a Moral Idea* (Oxford University Press, 2008). See also H.L.A. Hart, *The Concept of Law* (Clarendon Press, 2nd ed, 1994); Ronald Dworkin, *Law's Empire* (Harvard University Press, 1986); Oliver Wendell Holmes Jr., 'The Path of Law' (1897) 10 *Harvard Law Review* 457, 461; Lon L. Fuller, *The Morality of Law* (Yale University Press, 1969 ed) 106.

[25] Alec Stone Sweet, *Governing with Judges: Constitutional Politics in Europe* (Oxford, 2000) 11.

[26] Joe McIntyre, *The Judicial Function: Fundamental Principles of Contemporary Judging* (Springer, 2019) 13

[27] Ibid, 123, cf Dworkin Ronald Dworkin, *Law's Empire* (Harvard University Press, 1986) viii-ix

[28] H.L.A. Hart, *The Concept of Law* (Clarendon Press, second edition, 1994) 123, 128-36.

[29] Joe McIntyre and Jacqueline Charles, *Submission No 92 to Inquiry into Civics Education, Engagement, and Participation in Australia,* Joint Standing Committee on Electoral Matters, 29 May 2024; Joe McIntyre, The Ur-Controversy of Civil Courts (2022) in Marg Camilleri and Alistair Harkness (eds) *Australian Courts: Controversies, Challenges and Change* (Springer, 1st ed, 2022) 345.

[30] Luuk de Boer, 'Limit Cases: Sovereign Citizens and a Jurisprudence of Consequences' in M Hertogh and P de Winter (eds) *Empirical Perspectives on the Effects of Law: Towards Jurisprudence and Consequences* <https://figshare.com/articles/journal_contribution/de_Boer_L_O_forthcoming_b_Limit_Cases_Sovereign_Cit izens_and_a_Jurisprudence_of_Consequences_In_M_Hertogh_and_P_de_Winter_eds_i_Empirical_Perspectiv es_on_the_Effects_of_Law_Towards_Jurisprudence_of_Consequences_i_/28275701?file=51914321>.

The core argument of this paper, though, is directed to the nature of both phenomena (the use of LLMs and the use of pseudolaw). I argue that the apparent efficacy of both relies upon pareidolic illusions of meaning and that both mistake form for substance. To make out this argument, I provide a functional understanding of both phenomena. In Part II, I begin by explaining the nature of contemporary LLMs and provide a basic overview of how they produce outputs that appear to match human language and convey meaning. In Part III, I provide a brief overview of the phenomenon of pseudolaw and explain how adherents mistake the forms of legal ritual for law itself. In Part IV, I then argue that both phenomena share a common fault whereby users mistakenly perceive illusions as having substantive meaning

## PART II: THE NATURE OF LANGUAGE MODELS AS PSEUDO-LANGUAGE

By the start of 2025, most people have probably heard of Large Language Models ('LLMs')[31] like *ChatGPT*, *Gemini*, *Claude* etc, and have probably had a chance to experiment with the programs. The ability of LLMs to produce written text that look meaningful is 'extraordinary'.[32] As Kick notes, despite 'being merely complicated bits of software,' these models 'are surprisingly human-like when discussing a wide variety of topics.'[33]

Yet despite their impressive *looking* outputs, it is critical to appreciate, as I expand upon in this article, that this technology inherently content blind. Despite appearances, these programs do not undertake 'reasoning' in any recognisable form nor seek to perform 'natural language understanding' (NLU).[34] Rather, they are designed to process words-as-numbers through the probabilistic and statistical analysis of the relationships between incomprehensibly large sets of words-as-numbers. The output of this process *looks like language* in the way we understand that concept, but that is a chimeric illusion. As Floridi observes:

> [LLMs] … do not think, reason or understand; they are not a step towards any sci-fi AI; and they have nothing to do with the cognitive processes present in the animal world and, above all, in the human brain and mind, to manage semantic contents successfully. However, with the staggering growth of available data, quantity and speed of calculation, and ever-better algorithms, they can do statistically— that is, working on the formal structure, and not on the meaning of the texts they process —what we do semantically[35]

This is the key point to which I return to again and again in this article. LLMs do not reason, do not seek to verify or understand knowledge and develop an appreciation of the meaning of words and concepts. Rather, they are models are trained to 'predict the next word' through statistical analysis of huge datasets of text. As Choi *et al* note, LLMs are 'autoregressive', meaning that 'they predict the next word given a body of text' and then 'then repeatedly predict subsequent words to

---

compose indefinitely long bodies of text."[36] These models are *extraordinarily* good at this, so much so that it often seems that the programs understanding the underlying meaning, and externally verifying content. Yet this is chimeric. These models are trained on mindbogglingly large datasets, and assess billions of permutations for every prediction, to produce outputs that share the form of language but not its content and functional purpose.

However, to understand the nature of LLMs, and to properly appreciate the implications of this illusion, it is necessary to step beyond this explanation of the technology as simply some fancy autocomplete. Unfortunately, though, this is often where explanations about the operation of the technology stops, with the details of how these technologies work 'often treated as a deep mystery'.[37]

Partially, this is a function of the technology itself: LLMs like ChatGPT are programmed to utilise 'neural network' technologies,[38] a form of machine learning technology that allows models to process huge sets of training data to 'learn' and improve their accuracy over time.[39] In contrast to conventional software, where human programmers give explicit step-by-step instructions, neural networks are programmed to develop their own systems of operation. This is a form of 'black box' technology, whereby an output can be produced from a given set of inputs, but such models do not, and cannot, explain why they reach such conclusions. As currently designed it is functionally impossible to fully understand how these models operate as the precise operations of the neural networks are not programmed by humans. While research is being undertaken to better understand these models,[40] there are fundamental limits as to the extent to which these current models can be fully explained.

However, this should not be taken too far. The fundamental nature of what these models do, and how they operate, can be understood and explained. For the purposes of this article, it is not necessary to understand many of the processes by which LLMs, through 'transformers'[41] process text.[42] It is sufficient to grasp two key ideas:

1. Words are processes through a mathematical statistical process whereby each word is reduced to a complex multivariant set of numbers ('word vectors'); and
2. These models treat *words*,[43] rather than entire sentences or passages, as the basic unit of analysis

---

[36] Jonathan H. Choi et al, 'ChatGPT Goes to Law School' (2022) 71 *Journal of Legal Education* 387.The authors give the following example: 'given the phrase "I walked to the", a GPT model might predict that the next word is "park" with 5% probability, "store" with 4% probability, etc.' at 387.

[37] Timothy B. Lee and Sean Trott, 'A Jargon-Free Explanation of How AI Large Language Models Work', *ARS Technica* (online, 31 July 2023) <https://arstechnica.com/science/2023/07/a-jargon-free-explanation-of-how-ai-large-language-models-work/>.

[38] Ashish Vaswani et al, 'Attention is All you Need' (Conference Paper, 31st Conference on Neural Information Processing Systems, 2017).

[39] 'What is a Neural Network?', *IBM* (Web Page) <https://www.ibm.com/think/topics/neural-networks>.

[40] Kevin Wang et al, 'Interpretability in the Wild: A Circuit for Indirect Object Identification in GPT-2 Small', *arXiv* (online, 1 November 2022) <https://arxiv.org/abs/2211.00593>.

[41] Ashish Vaswani et al, 'Attention is all you need' (Conference Paper, 31st Conference on Neural Information Processing Systems, 2017).

[42] For an excellent and accessible explanation of how LLMs operate, see Timothy B. Lee and Sean Trott, 'A jargon-free explanation of how AI large language models work', *ARS Technica* (online, 31 July 2023) <https://arstechnica.com/science/2023/07/a-jargon-free-explanation-of-how-ai-large-language-models-work/>.

[43] Technically, these programs focus on 'tokens' rather than entire words, but for the purposes of clarity and accessibility, this article will refer to 'words' as the basic unit of analysis.

In the following sections I will outline the basic parameters of how LLMs use these ideas to construct sentences that appear to contain meaning.

However, before we begin this process, we must understand something about how humans use language, and the role 'meaning' and its transfer play in that process.

## 1. Syntax, Semantics and Intentionality

Language as a mode of communication is, fundamentally, a shared endeavour that depends upon not on the specific words themselves, but on their capacity to transfer meaning about shared concepts and experiences. The alternative – that words are self-defining and not communal – is absurd, as Lewis Carroll memorably illustrated in *Alice through the Looking Glass*: [44]

> 'When I use a word', Humpty Dumpty said, in rather a scornful tone, 'it means just what I choose it to mean neither more nor less.'
> 'The question is' said Alice, 'whether you can make words mean so many different things.'
> 'The question is,' said Humpty Dumpty, 'which is to be master – that's all.'

As this Carroll example helps illustrate, when humans use words, we use them to communicate about *ideas*, *objects*, *things* that have an existence independently of the words. When I say, 'Paris is the capital of France,' I understand that there exists a country 'France' and a city 'Paris' that have some physical and conceptual existence independently of that sentence. The words are a mode of communicating about those preexisting entities. It is not that those words have some preexisting absolute meaning; rather those words have a contextual and purposive meaning, a heady mix of intention, experience and shared community.

At its core, language is a collective enterprise, with the 'meaning' of any word dependant upon the way it is used a received by members of a community. As Kroeger notes,

> Humpty Dumpty's claim to be the "master" of his words – to be able to use words with whatever meaning he chooses to assign them – is funny because it is absurd. If people really talked that way, communication would be impossible.[45]

While language does, of course, exist independently of individuals, it cannot exit independently of community.[46] This highlights perhaps the 'most important fact' about word meaning and language: word meaning must be shared by the speech community.[47] For language to be functional, speakers 'must agree, at least most of the time, about what each word means.'[48] Saussure provides a useful metaphor when he argues that language 'exists in the form of a sum of impressions deposited in the brain of each member of a community, rather like a dictionary of which identical

---

copies have been distributed to each individual and that it exists.'[49] In this way, language is 'something that is in each [member of the community], while at the same time common.'[50]

It follows that to understand how humans use language – a necessary precursor to understanding the efficacy of LLMs use of 'language' – it is necessary to understand underlying ideas of meaningfulness. There are, of course, many ways to understand 'meaning'. The three principal disciplines concerned with the systematic study of 'meaning' are psychology, philosophy and linguistics.[51] For this paper, my focus is on the third of these: the linguistic discipline of semantics. Semantics is the systematic study of meaning,[52] or, more accurately, the study of the relationship between linguistic form and meaning.[53] The object of semantics is to understand how languages organize and express meanings.[54] As Kroeger explains:

> Linguists want to understand how language works. Just what common knowledge do two people possess when they share a language— English, Swahili, Korean or whatever—that makes it possible for them to give and get information, to express their feelings and their intentions to one another, and to be understood with a fair degree of success?[55]

Semantics can be contrasted with *syntax*, the study of the rules and practices by which words are arranged into phrases and sentences, (involving concepts such as word order, grammar and sentence structure),[56] and with *pragmatics*, the study of those aspects of meaning that depend on the way in which the words and sentences are used (including implicature, speech acts and non-verbal communication).[57] While these are each seen as distinct linguistic disciplines, each is obviously related: as Kreidler notes 'the meaning of a sentence is more than the meanings of the words it contains, and the meaning of a word often depends partly on the company it keeps.'[58] Each of these fields are substantial in their own right, and it is not my purpose to attempt to summarise even the foundational key concepts here. Rather, it is sufficient to note that there is a substantial scholarship about 'meaning' developed in each of these fields, and a clear recognition that (a) language exists to convey meaning and information between member of a group; and (b) there must be a sufficient degree of common understanding of meaning to enable that language to operate.

Moreover, and critically for the context of LLMs, linguistics – and semantics in particular – recognises that language is fundamentally a product of the human mind. This is particularly apparent in the way semantics is seen to possess both external and internal aspects: the former that examines how words refer to objects in the world (including what makes a sentence 'true') while the latter is interested in the connection between words and the mental phenomena they evoke

---

related.[59] Language is use to express meaning to others, to transfer ideas from one mind to another mind. To adopt the Kantian idea, whereas in philosophy we distinguish between the 'noumenon' (the thing-in-itself that exists independently of human sense) and the 'phenomenon' (the object as it appears in human experience), in semantics language allows us to transfer information about one person's phenomenon to affect the experience/phenomenon of another person. Language may be used to express meaning in a manner understandable by others, critically allowing us to express meaning that exists in our minds in a manner that is comprehensible to others.[60]

This is not a passive activity utilising predetermined and substantively complete word-codes. Rather, the meaning of words is fundamentally integrated with, and dependent upon, the intentions of the users. As Reimer notes, we 'do not just use language to talk about or describe the world'; rather, we use language 'in order to manipulate and induce transformations in it.'[61] Indeed, for Givon, the 'ultimate purpose' of any communicative transaction is 'to manipulate the other toward some target action.'[62] It follows that while we might passively say that language draws upon an underlying shared body of knowledge to convey meaning, the communicative purpose in the use of language inevitably shapes that meaning. This recognised in the of three different types of linguistic units of 'meaning' Kroeger discusses, namely:

1. Word meaning
2. Sentence meaning
3. Utterance meaning (also referred to as "speaker meaning")[63]

While the first two types of meaning are relatively self-explanatory, the third category is worth pausing upon. In the context of semantics, a 'sentence' is a linguistic expression (a well-formed string of words) while an utterance is a 'speech event' by a particular speaker in a specific context.[64] When a speaker uses a sentence in a specific context, with a particular intentionality, the speaker produces an utterance. Whereas the 'sentence meaning' refers to the semantic meaning which derives from the words themselves, regardless of context,[65] the utterance meaning builds upon this to include any pragmatic meaning intended by the speaker. In this way Cruse defines utterance meaning as 'the totality of what the speaker intends to convey by making an utterance.'[66]

Taken as whole, the study of meaning though semantics highlights that language is a deliberate communicative practice that operates to alter world by transferring meaning from one mind to another. The discipline analyses the ways in which words and language structures operate to create and convey meaning, highlighting the communitive, communicative and intentional qualities of language.

## 2.  Words-as-Numbers: Vectors, Transformers and Statistical Modelling

LLMs use and manipulate words in a fundamentally different manner to this semantically driven model of human language. Most apparently, all meaning resides with the recipient of the 'language output': LLMs are not members of a language community and lack any form of mind capable of experiencing the phenomenon of any noumenon, let alone engaging in an information transfer from one mind to another. More specifically, such models lack agency and intentionality in the sense Givon notes, as there cannot be any intention to impact upon the world without some form of consciousness. LLMs necessarily lack any internal semantics. Yet these models also exclude any form of 'utterance meaning', as there is no context or pragmatic meaning to contribute to formal meaning. Further, as I outline below, by using the word as the unit of analysis, these models do not engage in any form of 'sentence meaning'. There is no attempt to develop any syntax, with no modelling of rule of grammar or sentence composition.

Yet somehow, the models can produce outputs that seem indistinguishable from meaningful language.

To understand how this occurs it is necessary to understand some basics of mathematical concepts such as 'vectors' and matrix multiplication and how these can be used to represent and manipulate 'words-as-numbers'. When humans construct a representation of a word in English we use a sequence of letters – say O-C-E-A-N for 'ocean', and then rely upon shared knowledge of that word to convey meaning (eg its wet, fishy and large…). LLMs instead use a long list of numbers called a 'word vector'[67] to both represent a word and to encode 'meaning'.

### (a)  Words as Vectors – Numbers as Meaning

In mathematics, a 'vector' is an object that has both a magnitude and a direction. Geometrically, we can picture a vector as a directed line segment, whose length is the magnitude of the vector and with an arrow indicating the direction (Figure 1). We can use vector notation to represent concepts where both magnitude and a direction matter – such as when dealing with forces in physics or engineering. What is particularly useful in utilising vectors is that we can represent such a directed line segment on a Euclidean plane, which enables mathematical representation and manipulation (Figure 2):

---

**Figure 1: Representation of a Vector**

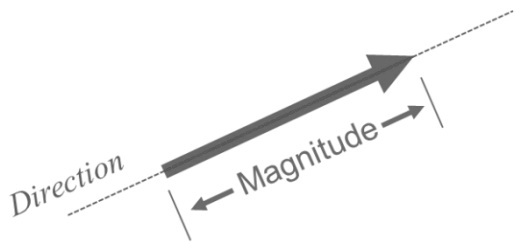

**Figure 2: Vector on a Euclidean Plane**

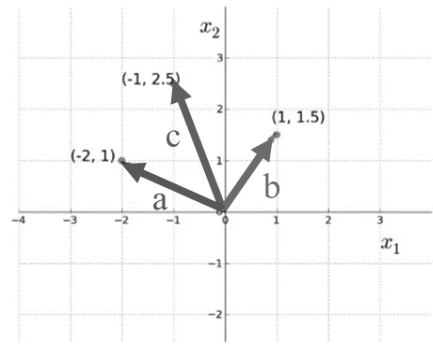

In the above Figure 2, we can represent the vectors a and b in matrix notation |**a**| and |**b**| such that if:

$$\text{If} \qquad a=\begin{bmatrix} -2 \\ 1 \end{bmatrix} \quad \& \quad b=\begin{bmatrix} 1 \\ 1.5 \end{bmatrix} \qquad \text{then} \qquad a + b = c \qquad \begin{bmatrix} -2 \\ 1 \end{bmatrix} + \begin{bmatrix} 1 \\ 1.5 \end{bmatrix} = \begin{bmatrix} -1 \\ -2.5 \end{bmatrix}$$

What is particularly useful here is that while it is visually impossible to represent vectors in more than three dimensions, mathematically, it is straightforward to use and manipulate matrices that involve very large sets of variables.

Representation of words as continuous vectors has a long history,[68] but the concept was thrust into the mainstream when Google announced its 'word2Vec' project in 2013.[69] That project involved the development of 'two novel model architectures for computing continuous vector representations of words from very large data sets.'[70] In that case, the training dataset involved 6 billion 'tokens' drawn from the Google News corpus,[71] generally using 640 dimensional word vectors.[72] That is, each word was represented by a vector of 640 discrete numbers, which allowed the capture of relationships between word-vectors.[73] Prior models had databases of millions of words with 10s of dimension word vectors, with early models relying upon human labelling. The word2vec project was revolutionary in two ways: firstly, it illustrated the power of neural network machine learning to label continuous vector representations of words from very large data sets, and secondly it demonstrated the utility of such labelling. As the authors state:

---

Using a word offset technique where simple algebraic operations are performed on the word vectors, it was shown for example, that *vector("King") - vector("Man") + vector("Woman")* results in a vector that is closest to the vector representation of the word *Queen* [74]

Effectively, these high-dimension word vectors allowed a degree of 'reasoning' about words: by manipulating the underlying matrices, probabilistic analysis could reveal relationships. Such relationships could be defined by subtracting two-word vectors, with the result is added to another word. [75] For example, (*vector('Paris') – vector('France')) + vector('Italy') = vector('Rome')*. Effectively, word vectors allow models to encode 'subtle but important information about the relationships between words.'[76]

This approach of words-as-number vectors (or 'word embeddings') has since become a foundational architecture of NLP systems. These high-dimensional vector space representations are based on the 'distributional hypothesis' that words with a similar distribution have similar meanings,[77] essentially suggesting that 'the meaning of the word can be induced from a large number of texts.'[78]

It is worth pausing on this idea.

The central hypothesis of contemporary LLMs is that semantic meaning can be derived indirectly and by proxy by examining lexical terms with a similar distribution in a sufficiently large corpus of texts. Essentially, this supposes that sufficiently high-dimensional word-vectors derived from sufficiently large datasets will embed sufficient information to allow meaning to be mathematically inferred. Words are reduced to numbers and matched against other words-as-numbers, with sufficiently similar words treated as having a shared meaning. In such models, there is no attempt to verify abstract meaning, to ensure that there is any substantive conformity between the modelled meaning and the use of that word in 'real world' linguistic community. It is numbers all the way down.

A visual example can be useful here. The *WebVector* project[79] based on research undertaken be Fares *et al*,[80] allows the illustration of how word vectors operate. In the example below the noun

---

'ocean' is represented by a word vector of 300 dimensions (the first three values are 0.0045118448324501514, -0.12667645514011383, 0.4031252861022949) derived from the corpus of English *Wikipedia*. The figures below illustrate word vectors with similar distributions:

**Figure 1: Vectorial Representation of the word 'ocean'[81]**

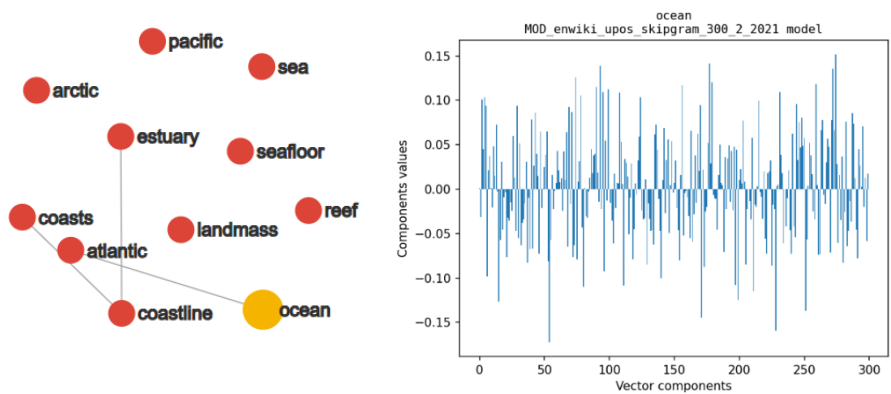

The idea is that by increasing dimensionality and the size of the training set, accuracy of meaning can be improved. For example, GPT1 had 768 dimensional vectors, GPT2 had 1600 dimensional vectors and GPT3 had 12,288 dimensional vectors.[82]

Of course, there are many limitations with this approach. Firstly, these vectors are limited by the scope and content of the underlying training dataset. As Sedinkina notes the word 'god' will be seen to have entirely different meanings dependent upon whether it is trained upon a financial corpus or a general corpus. The figure below illustrates different neighbours depending upon that training corpus:

**Figure 2: Vectorial Representation of 'God' by Different Training Corpus[83]**

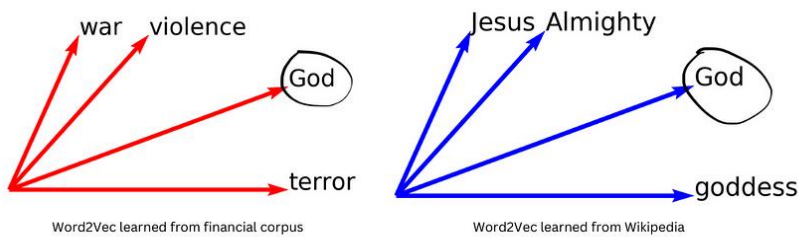

These different corpora produce word vectors with completely different meanings. While of course, the meaning of words will reflect context, this illustration highlights the limitation of the distribution hypothesis as the sole meaning of determining 'meaning'. One response to this issue is simply to increase the dimensionality of word vectors. However, this solution bleeds into the second challenge, namely the 'curse of dimensionality',[84] whereby the sheer number of variables

---

that can be collected can be problematic.[85] As Wadkar notes, in 'high-dimensional vector spaces … randomly selected vectors tend to be *equidistant* from one another', with the result that common distance metrics break down and become less meaningful.[86] Essentially, the noise-to-signal ratio becomes increasingly problematic, and while workarounds – like 'clustering' that focuses only on a small subset of dimension – assist,[87] the problem remains.

Thirdly, as Lee and Trott note, because these vectors 'are built from the way humans use words, they end up reflecting many of the biases that are present in human language',[88] such as occupational gender biases (eg (*vector('doctor') – vector(man)*) *+ vector(woman) = vector(nurse)*).[89]

Nevertheless, LLMs that are trained on large datasets (GPT-3 was trained on a corpus of approximately 500 billion words) have shown a remarkable capacity to use high-dimensional word vectors to encode information about words and their relationship to each other as they appear in various textual corpuses.

Whether this amount to a capacity to encode semantic meaning is something I will return to below. What matters here is that LLMs code 'meaning' as something internal to words, any external validation or verification of that meaning.

### *(b) Transformers and Attention: One Word at a Time*

First, however, it is necessary to understand something about the second key aspect of contemporary LLM architecture and how it enables these models to treat words, rather than entire sentences or passages, as the basic unit of analysis.

The meaning of words in natural language depends upon context,[90] and words often have multiple meanings. Given the above analysis, it should not be surprising that LLMs can 'represent the same word with different vectors depending on the context in which that word appears',[91] with more similar vectors for more closely related meanings.[92]

For humans, we can distinguish between words by considering external context and by understanding facts about the world: Jill may fossick at the [river] bank but will meet her advisor at the [financial] bank. LLMs do not seek external context or verification to determine meaning. So how is context assessed?

---

[85] Naomi Altman and Martin Krzywinski, 'The Curse(s) of Dimensionality' (2018) 15 *Nature Methods* 399.

[86] Sameer Wadkar, 'The Curse of Dimensionality, 768-Deminsional Word Embeddings, and LLM Perplexity – How These Concepts Interact', *LinkedIn* (online, 30 September 2024) <https://www.linkedin.com/pulse/curse-dimensionality-768-dimensional-word-embeddings-llm-wadkar-h5lde/>.

[87] See Ira Assent, 'Clustering High Dimensional Data' 2(4) *WIRES Data Mining and Knowledge Discovery* 340.

[88] Timothy B. Lee and Sean Trott, 'A Jargon-Free Explanation of How AI Large Language Models Work', *ARS Technica* (online, 31 July 2023) <https://arstechnica.com/science/2023/07/a-jargon-free-explanation-of-how-ai-large-language-models-work/>; See https://www.science.org/doi/full/10.1126/science.aal4230

[89] There are a range of other concerns, including the tendency of training data to encode hegemonic worldviews: Emily M. Bender et al, 'On the Dangers of Stochastic Parrots: Can Language Models Be Too Big?' (Conference Paper, Conference on Fairness, Accountability, and Transparency, 3-10 March 2021) 616.

[90] See Charles W. Kreidler, *Introducing English Semantics* (Routledge, 1998) 8.

[91] Timothy B. Lee and Sean Trott, 'A Jargon-Free Explanation of How AI Large Language Models Work', *ARS Technica* (online, 31 July 2023) <https://arstechnica.com/science/2023/07/a-jargon-free-explanation-of-how-ai-large-language-models-work/>.

[92] Vinod Nair and Geoffrey E Hinton, 'Rectified Linear Units Improve Restricted Boltzmann Machines' (Conference Paper, 27th International Conference on Machine Learning, 21 June 2010) 807.

Again, the key breakthrough was a paper produced by Google.[93] In 2017, *Transformer* architecture was released[94] which allowed autoregressive (predict the next word only) generative models based on neural network 'feed forward' architecture. The key idea 'self-attention' approach is that each layer of this model allows the context of the focal word to be assessed by reference to other inputs in the sequence, and to 'feed forward' what is learnt in the first layer to subsequent layers. As Sedinkina describes:

> A multi-head self-attention mechanism is the main and the most important component in this architecture — as the model processes each word, self-attention looks at other positions in the input sequence, finds the relevant words and uses this information to encode the currently processing word. If we consider the sentence *"The cat sat on a mat because it was too tired."*, we will associate *"it"* with *"cat"* and not with *"mat"*. This is exactly what self-attention does — it allows to associate *"it"* with *"cat"*, when the model is processing the word *"it"*.[95]

Modern LLMs utilise significant numbers of layers (GPT-3 has 96 layers), with research suggesting that initial layers focus on syntax and ambiguities and subsequent layers on higher-level understanding.[96] The final layer will, as Lee and Trott note, 'output a hidden state for the final word that includes all of the information necessary to predict the next word.'[97]

Again, it is difficult to comprehend the scale of the analysis that is occurring in this modelling. For example, in GPT-3, there are 49,152 neurons in the hidden layer (each with 12,288 inputs), and 12,288 output neurons (each with 49,152 input values) for 1.2 billion parameters. Over the 96 feed-forward layers, there is a total of *116 billion* parameters *for each word output*.[98] Through the analysis of each feed-forward layer, which examines only one word at a time, this transformer process allows the other word vectors in each input to provide contextual information for the focal word. While the numbers of parameters being processed are astronomically large it is implemented as a chain of simple mathematical functions, which enables them to 'learn' by trying to predict the next word in ordinary passages of text.[99] In this manner 'context' can be assessed from within the universe of word vectors, drawing upon patterns identified in the training data. Again, there is no – and no attempt to provide – external validation or verification of context.

### 3.   LLM as Pseudo-language

While there are many other brilliant operations involved, foundational architecture of contemporary LLMs is based on these two features: (1) the use of word vectors to code information about words, and (2) the use of transformers to allow models to focus on words as the unit of analysis by drawing on context from other input words. The basic matrix manipulation involved

---

in the processes of these word-vectors is relatively straightforward. Yet the scale involved is truly mind-boggling: GPT-4 was trained on a corpus of 500 billion words, uses 12,288 dimensions for each word-vector, uses 49,152 neurons in each feed forward layer to generate 116 billion parameters *for each word processed*. These numbers are effectively impossible to comprehend. But the final output is stunning: as Floridi note, these models 'process text with extraordinary success and often with outcomes that are indistinguishable from those that human beings could produce.'[100]

Of course, this apparently stunning validation of the Distribution Hypothesis comes at a cost: these models consume extraordinarily large amounts of energy[101] and water[102] to operate, are incredibly expensive to build and run,[103] and yet do not generate significant income.[104] It is reported that advances in the technology are beginning to plateau,[105] and despite the eye-watering investments, no one has been able to yet produce a 'meaningful, industry-defining' application.[106] And underlying all this is the issue of the unreliability of the outputs, in terms of the 'actual' real meaning of words, phrases and assertions. This issue is commonly referred to as 'hallucinations' and arises when an LLM has 'plausible yet nonfactual content'[107]. Again, Floridi usefully highlights the inherent fragility of these models in the following passage:

> They are fragile, because when they do not work, they fail catastrophically, in the etymological sense of a vertical and immediate fall in the performance. … They make up texts, answers or references when they do not know how to reply; make obvious factual mistakes; sometimes cannot make the most trivial logical inferences or struggle with simple mathematics including the numbers in crochet instructions; or have strange linguistic blind spots where they get stuck.'[108]

---

[100] Luciano Floridi, 'AI as *Agency Without Intelligence*: on ChatGPT, Large Language Models, and Other Generative Models' (2023) 36, 15 *Philosophy and Technology* 1, 2.

[101] As Swan notes, 'Generative AI already uses as much energy as a small country and is predicted to rival that of Japan within a year. Such searches use 10 times the energy of a normal web search, and the technology has tripled the energy requirements of the entire tech sector in just two years.': David Swan, 'Every time you use ChatGPT, half a litre of water goes to waste', *The Sydney Morning Herald* (online at 2 December 2024) <https://www.smh.com.au/technology/every-time-you-use-chatgpt-half-a-cup-of-water-goes-to-waste-20241128-p5kubq.html>; see Sourabh Mehta, 'How Much Energy Do LLMs Consume? Unveiling the Power Behind AI', *ADaSci* (online at 3 July 2024) <https://adasci.org/how-much-energy-do-llms-consume-unveiling-the-power-behind-ai/?ts=1737504745>; Radosvet Desislavov, Fernando Martínez-Plumed and José Hernández-Orallo, 'Trends in AI Inference Energy Consumption: Beyond the Performance-vs-Parameter Laws of Deep Learning' (2023) 38 *Sustainable Computing: Informatics and Systems* 100857.

[102] Matt O'Brien and Hannah Fingerhut, 'Artificial intelligence technology behind ChatGPT was built in Iowa – with a lot of water', *AP News* (online at 10 September 2023) <https://apnews.com/article/chatgpt-gpt4-iowa-ai-water-consumption-microsoft-f551fde98083d17a7e8d904f8be822c4>; Almando Morain et al, 'Artificial Intelligence for Water Consumption Assessment: State of the Art Review' (2024) 38(9) Water Resources Management 3113; A. Shaji George, A. S. Hovan George and A. S. Gabrio Martin, 'The Environmental Impact of AI: A Case Study of Water Consumption by Chat GPT' (2023) 1(2) *International Innovation Journal* 97-104 <https://puiij.com/index.php/research/article/view/39>.

[103] It is projected that $US200 billion dollars will be spent on AI in 2025: 'Big tech 2025 capex may hit $200 billion as gen-AI demand booms', *Bloomberg Professional Services* (online at 4 October 2024) <https://www.bloomberg.com/professional/insights/technology/big-tech-2025-capex-may-hit-200-billion-as-gen-ai-demand-booms/?ref=wheresyoured.at>.

[104] For example, OpenAI (the owner of the flagship ChatGPT), is expected to lose US$4-5 billion in 2024 alone: Edward Zitron, 'OpenAI Is A Bad Business', *Where's Your Ed At* (online at 2 October 2024) <https://www.wheresyoured.at/oai-business/>.

[105] Rachel Metz, Shirin Ghaffary, Dina Bass, and Julia Love 'OpenAI, Google and Anthropic Are Struggling to Build More Advanced AI', *Bloomberg.com* (online, 13 November 2024) <https://www.bloomberg.com/news/articles/2024-11-13/openai-google-and-anthropic-are-struggling-to-build-more-advanced-ai>.

[106] Edward Zitron, 'Godot Isn't Making It', *Where's Your Ed At* (online at 3 December 2024) <https://www.wheresyoured.at/godot-isnt-making-it/>.

[107] Lei Huang et al, 'A Survey on Hallucination in Large Language Models: Principles, Taxonomy, Challenges, and Open Questions' (2025) 43(2) ACM Transactions on Information Systems 1, 1.

Any use of LLMs is familiar with hallucinations. But what is rarely appreciated is that these are not a 'bug' in the system. It is not that the models normally get it right, but sometimes they get it wrong. Rather, hallucinations are inevitable and ever-present: everything that is produced by LLMs is an illusion of meaning, but most times, humans mistake that illusion for the real thing. A 'hallucination' is simply an output that the reader knows to be wrong.

### (c) The Inevitability of Hallucinations

The core problem here is that the architecture described above is designed to assess probabilistic relationships between word vectors, and not to assess meaning directly. Again, to repeat Bender's observation, these LLMs are 'not performing natural language understanding',[109] nor are they even attempting to do so. The models, instead, operate on two foundational assumptions:

(1) **The Correlation of Frequency and Meaning Inference:** Firstly, these models assume that the more common a correlation between words vectors in the training data, the stronger the inference that those word vectors share a common meaning. 'Meaning' is in this context is a property entirely internal to the training data and bears no direct relationship to 'meaning' as conventionally understood; and

(2) **The Convergence of External and Internal Meaning Inference:** Secondly, and drawing on the Distribution Hypothesis, these models assume that given a sufficiently large training data set, the ascribed meaning of a word vector will converge with the real world meaning.

At a sufficiently high level of generality, these assumptions seem pretty good. Most outputs appear to bear a strong relationship to the meaning and substance of words as used in natural language. The core problem, however, is that the more granular the focus the weaker the inferences. At that is because, at their core, these models depend upon *probability* rather than *intentionality*. Humans use language is inconsistent, creative and varied ways, in good faith and in bad, embracing both 'truth' and 'deception'. Probabilistic analysis may reveal what is most likely given the underlying corpus, but that says *nothing* about what is actually intended in a given usage. Hallucinations arise, in part, because a specific usage is at variance with the statistically most common usage. Consider the following example:

---

[108] Luciano Floridi, 'AI as *Agency Without Intelligence*: on ChatGPT, Large Language Models, and Other Generative Models' (2023) 36, 15 *Philosophy and Technology* 1, 3. For relevant research into the weaknesses of LLMs, Konstantine Arkoudas, 'GPT-4 Can't Reason', *ArXiv* (online, 10 August 2023) <https://arxiv.org/abs/2308.03762>; Tom Kocmi and Christian Federmann, 'GEMBA-MQM: Detecting Translation Quality Error Spans with GPT-4', *ArXiv* (online, 21 October 2023) <https://arxiv.org/abs/2310.13988>; Luciano Floridi and Massimo Chiriatti, 'GPT-3: Its Nature, Scope, Limits, and Consequences' (2020) 30 *Minds and Machines* 681; Arush Tagade and Jessica Rumbelow, 'Prototype Generation: Robust Feature Visualisation for Data Independent Interpretability', *ArXiv* (online, 29 September 2023) <https://arxiv.org/abs/2309.17144>; Karl Cobbe et al, 'Training Verifiers to Solve Math Word Problems', *ArXiv* (online, 18 November 2021) <https://arxiv.org/abs/2110.14168>; Ethan Perez et al, 'Discovering Language Model Behaviors with Model-Written Evaluations', *ArXiv* (online, 19 December 2022) <https://arxiv.org/abs/2212.09251>.

[109] Emily M. Bender et al, 'On the Dangers of Stochastic Parrots: Can Language Models Be Too Big?' (Conference Paper, Conference on Fairness, Accountability, and Transparency, 3-10 March 2021) 610.

**Figure 3: Illustration of ChatGPT Hallucination**[110]

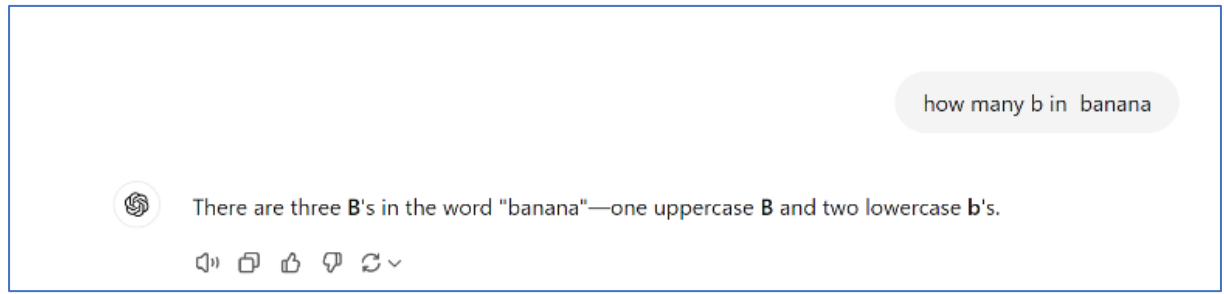

While it is not possible to directly determine the 'reasoning' behind the hallucination in this example, it is likely to be a relic of the way the program processes the 'tokens' that are represented in the word. Alternatively, the model may be associating this question with the more common pattern seen in the dataset (perhaps 'how many a in banana') and answering that question. While humans have no issue recognising that the mere frequency of a statement has no necessary correlation with the truth of that statement, such correlations are fundamental to the inferences of 'meaning' for LLMs. These models operate on the assumption that sheer scale will eliminate discordant instances – yet such instances may not only be the intended meaning but may be more truthful than the dominant usage.

Fundamentally, LLMs designed to operate on the basis that '*Meaning-as-Internal-Vector-Relationships*' will, over a large enough set, in some manner approximate '*Actual Meaning*.' However, there is no architectural design or infrastructure involved in verifying or validating this assumption. When I make assertions in this paper, I validate them by reference to external sources to ensure the consistency and coherence of my word usage, and quality and reliability of those assertions. In contrast, with LLMs, its '*Meaning-as-Internal-Vector-Relationships*' all the way down.

It follows that while the output of LLMs is very good at approximating the form of language, not only are they structurally blind to 'Actual Meaning', but they lack any intentionality or purpose in the transference of meaning. LLMs produce word-like outputs but are not in any meaningful way *communicating*. As seen in the first part of this section, linguistic semantics engages in a substantial inquiry to investigate meaning, drawing on issues of issues of semantic, sentence and utterances meanings to help understand how *this* person intended for *this* word to be understood in *this* context, and to understand the intended purpose of that transference of meaning. The frequency-based analysis of word-vector relationships in LLMs cannot fundamentally do this. Rather they work through the operation of probabilities, which is necessarily incapable of achieving certainty in any specific instance. At that level, at the point of intentionality of a given person at a given time, probabilistic analysis breaks down. Yet, this is the level at which humans use language.

It follows, that while these models can achieve a relatively high degree of *apparent* accuracy of meaning, they possess a completely unknown degree of *actual* accuracy with respect to external meaning. This makes these models inherently unreliable with regard to the substantive content of the output. If the user wishes to produce something that looks/appears/has the form of something meaningful, then LLMs are fantastic. Yet these models are entirely inadequate if the user wants something that is reliably and demonstrably accurate.

---

[110] ChatGPT, OpenAI, Correspondence with Joe McIntyre, 21 January 2025.

Ultimately, for LLMs, hallucinations *are not* a problem that can be sanded off or avoided through better design. Rather, hallucinations are baked in; no *mistake* in some instances, but the *entirety* of the output is wrong. Literally, the whole output is a hallucinated dream - what is impressive is how often it tricks us into thinking there is meaning. Effectively, a hallucination is simply an output we as users know is wrong, nothing more and nothing less.

By focusing entirely on the corpus of words, and making individual words the sole object of analysis, LLMs are using the form of language in a mode entirely different to the way in which humans use language to communicate. The objective of such models is to achieve predictions based upon relative relationships of word vectors, mathematically dealing with 'words-as-numbers' in a manner that is logical, probabilistic and deterministic. 'Actual Meaning' is, in this process, incidental and accidental, if not entirely irrelevant. There is no relationship to the hierarchically prior phenomenon. There is no role for transference of meaning between minds. 'Words' are being used in a fundamentally different way from how humans use them. This is not language but pseudo-language. There is no meaning, only the illusion of it.

What is, however, perhaps most significant in understanding the social phenomenon of the rise of LLMs is not that they produce a pseudo-language but that humans appear to have an irresistible tendency to treat it as real language and that it conveys genuine, accurate and reliable meaning.

### *(d) Syntax, Semantics and Stochastic Parrots*

In computer science, the relationships between the form and content of the use of language have been studied for many decades. Most notably, Searle demonstrated over 40 years ago, that the effective use of formal syntax cannot be equated with the generation of substantive meaning or understanding. This is particularly relevant in the context of contemporary LLMs that use of word vectors to develop a largely coherent emergent syntax and, in turn, to construct outputs that largely share the form of language. But in contrast to the normative and communal nature of human syntax, this is a formal and atomised syntax. Yet there appears to be a tendency to treat this as genuine syntax that reflects and engages with underlying semantics. This is, though, fundamentally wrong.

In his influential 'Chinese Room Argument',[111] first articulated in 1980, Searle demonstrated the insufficiency of formal syntactical language systems to generate genuine semantic meaning on their own. In a later article, Searles described the thought experiment as follows:

> Imagine a native English speaker who knows no Chinese locked in a room full of boxes of Chinese symbols (a data base) together with a book of instructions for manipulating the symbols (the program). Imagine that people outside the room send in other Chinese symbols which, unknown to the person in the room, are questions in Chinese (the input). And imagine that by following the instructions in the program the man in the room is able to pass out Chinese symbols which are correct answers to the questions (the output). The program enables the person in the room to pass the Turing Test for understanding Chinese but he does not understand a word of Chinese.[112]

---

[111] See John Searle, 'Minds, Brains and Programs' (1980) 3 *Behavioural and Brain Sciences* 417–57.

[112] Robert A. Wilson and Frank C. Keil (eds.), *The MIT Encyclopedia of the Cognitive Sciences* (The MIT Press, 1999) 115.

Searle argued that programs implemented by computers are just syntactical. Computer operations are "formal" in that they respond only to the physical form of the strings of symbols, not to the meaning of the symbols.[113] As such, these models cannot generate actual semantic meaning. In Searle's words:

> Computation is defined purely formally or syntactically, whereas minds have actual mental or semantic contents, and we cannot get from syntactical to the semantic just by having the syntactical operations and nothing else. [114]

The focus of this argument, and the substantial critique and engagement with it over the last five decades, has been on whether the model can be said to 'understand' the meaning of words used by it.[115] That is, the focus of this debate is of the generator of 'meaning', the 'speaker' in form of the LLM, and whether it

However, the true significance, for me, is found not in the generation of apparent meaning, but in its receipt. This issue is highlighted in the influential 2021 article of Bender,[116] where she outlined the real-world risk of harm that arises from 'the tendency of researchers and other people *to mistake LM-driven performance gains for actual natural language understanding.*'[117] In that article, Bender highlights that – contrary to the way in which language is used by humans- text generated by an LLM is 'not grounded in communicative intent, any model of the world, or any model of the reader's state of mind.'[118] Rather, these models are effectively mimicking the form of human language without any underlying substantive meaning. Bender illustrated this by introducing the metaphor of the 'stochastic parrot', arguing that an LLM is[119]:

> …a system for haphazardly stitching together sequences of linguistic forms it has observed in its vast training data, according to probabilistic information about how they combine, but without any reference to meaning: a stochastic parrot.[120]

Thought this metaphor has been the most celebrated aspect of Bender's article, for me the revolutionary insight is Bender's recognition that what is actually occurring is a discrete psychological phenomenon whereby humans demonstrate a tendency to mistake *not-language* and *pseudo-language* for language itself. After outlining the ways in which 'human-human communication is a jointly constructed activity' between persons who share ground and use language to convey communicative intent,[121] Bender observes that the text generated by LLMs is not 'grounded in communicative intent, any model of the world, or any model of the reader's state of mind'. [122] She continues, critically, to observe that:

> This can seem counter-intuitive given the increasingly fluent qualities of automatically generated text, but we have to account for the fact that *our perception of natural language text, regardless of how it was generated, is mediated by our own linguistic competence and our predisposition to interpret communicative acts as conveying coherent meaning and intent, whether or not they do*. … The problem is, if one side of the communication does not have meaning, then the *comprehension of the implicit meaning is an illusion* arising from our singular human understanding of language (independent of the model). [123]

---

[113] 'The Chinese Room Argument', *Stanford Encyclopaedia of Philosophy* (online at 19 March 2004) 5.1 <https://plato.stanford.edu/entries/chinese-room/#SyntSema>.

[114] John Searle, 'Why Dualism (and Materialism) Fail to Account for Consciousness', in Richard E Lee (ed.), *Questioning Nineteenth Century Assumptions About Knowledge, Iii: Dualism* (Suny Press, 2010) 5.

[115] For a useful overview of the argument and its receipt, see: 'The Chinese Room Argument', *Stanford Encyclopaedia of Philosophy* (online at 19 March 2004) <https://plato.stanford.edu/entries/chinese-room/>.

[116] Emily M. Bender et al, 'On the Dangers of Stochastic Parrots: Can Language Models Be Too Big?' (Conference Paper, Conference on Fairness, Accountability, and Transparency, 3-10 March 2021).

[117] Emily M. Bender et al, 'On the Dangers of Stochastic Parrots: Can Language Models Be Too Big?' (Conference Paper, Conference on Fairness, Accountability, and Transparency, 3-10 March 2021) 616 (emphasis added).

[118] Emily M. Bender et al, 'On the Dangers of Stochastic Parrots: Can Language Models Be Too Big?' (Conference Paper, Conference on Fairness, Accountability, and Transparency, 3-10 March 2021) 616.

[119] The Oxford English Dictionary defines *stochastic* as 'randomly determined; that follows some random probability distribution or pattern, so that its behaviour may be analysed statistically but not predicted precisely': 'Stochastic', *Oxford English Dictionary* (online at 14 March 2025) [def 2a].

[120] Emily M. Bender et al, 'On the Dangers of Stochastic Parrots: Can Language Models Be Too Big?' (Conference Paper, Conference on Fairness, Accountability, and Transparency, 3-10 March 2021) 617.

[121] Bender observes: Our human understanding of coherence derives from our ability to recognize interlocutors beliefs and intentions within context. That is, human language use takes place between individuals who share common

This is, perhaps, the most important thing to appreciate in any analysis of the social use of LLMs. And in many respects, this insight is overshadowed by the image of the stochastic parrot. Because while the parrot reflects Searle focus on the 'speaker' of the LLM, here Bender highlights the root cause of the problem – namely the human *'predisposition to interpret communicative acts as conveying coherent meaning'.* When we consider the stochastic parrot, the focus is on the LLM and its ability to create intention free pseudo-meaning. While this is accurate, and an important observation, it underplays the real danger and nature of the phenomenon that arises from the behaviour of the 'recipient' (the human reader) who ascribes patterns and meaning where there are none.

Rather, as I will elaborate on below, LLMs take advantage of the pareidolic tendencies of the human mind: these models create images and illusions of meaning that the human users mistakenly construct into a meaningful pattern and erroneously ascribes meaning. The most important part of this phenomenon in not the output of the LLM, but the receipt of it by the human user: an illusion of meaning is created, and we humans ascribing to it a substance it does not possess.

## PART III: SOVEREIGN CITIZENS, PSEUDOLAW AND THE USE OF PSEUDO-LEGALISM

It may appear odd to transition at this point to a discussion of the emergent social phenomenon of pseudolaw. The relationship between the use of LLMs and of pseudolaw is not one is immediately apparent. Yet the central claim I develop above is as to the pareidolic illusion of meaning in the use of LLMs is similarly central to the use and 'success' of growing phenomenon of pseudolaw. As I outline below, pseudolaw makes use of many of the forms and sources of traditional law to create new rules that have the illusion of being legally meaningful, with adherents ascribing to it a substance it does not possess. This is essentially the precise same conceptual mechanism that underlies the misuse of LLMs. Both phenomena converge through their facilitation of 'conceptual pareidolia' – on the misguided construction of meaning from disparate forms. And it is the convergence itself that helps to illuminate this processing error in both discrete instances.

---

ground and are mutually aware of that sharing (and its extent), who have communicative intents which they use language to convey, and who model each other's mental states as they communicate. As such, human communication relies on the interpretation of implicit meaning conveyed between individuals. … *human-human communication is a jointly constructed activity*': Emily M. Bender et al, 'On the Dangers of Stochastic Parrots: Can Language Models Be Too Big?' (Conference Paper, Conference on Fairness, Accountability, and Transparency, 3-10 March 2021) 616.

[122] Emily M. Bender et al, 'On the Dangers of Stochastic Parrots: Can Language Models Be Too Big?' (Conference Paper, Conference on Fairness, Accountability, and Transparency, 3-10 March 2021) 616.

[123] Emily M. Bender et al, 'On the Dangers of Stochastic Parrots: Can Language Models Be Too Big?' (Conference Paper, Conference on Fairness, Accountability, and Transparency, 3-10 March 2021) 616.

To make this argument, however, it is necessary to understand the basic parameters of the emerging phenomenon of pseudolaw.

The term 'pseudolaw' is used to refer to the social phenomenon whereby adherents use the *forms* of conventional legal argument, but not the substance, in their interactions with law, legal and government institutions. As Mark Pitcavage, one of the preeminent global scholars on the topic, describes it, pseudolaw can be understood as:

> 'the creation and employment of theories and arguments that mimic the outward form and appearance of legitimate legal arguments to the extent that they might be able to convince an untrained layperson, but which in no meaningful way (except rarely and by accident) address or have standing in actual law[124]

The use of pseudolaw has been around as a distinct phenomenon for at least 50 years,[125] but has seen explosive growth since the COVID-19 global pandemic.[126] It has now reached the point where this 'unfortunately growing phenomenon'[127] is now 'clogging up' courts in many jurisdictions,[128] significantly increasing the workload of judges and administrators.[129]

## 1. The Rise of Pseudolaw

Of course, the use by individual litigants of nonsensical and ineffective arguments that mimic the appearance of law is not, of itself, a new form of behaviour. Indeed, as Pitcavage notes, the use of pseudolegal arguments by individuals has existed 'from the distant time the first misguided *pro se* litigant or defendant entered a courtroom to present a fantastical interpretation of the law that they had convinced themselves would extract them from whatever legal difficulties they faced.'[130]

What has changed is that whereas previously, these arguments were developed and used by specific individuals only to disappear with each act of litigation, what began to emerge in the second half of the 20th century was a range of semi-structured movements/groups using pseudolegal argumentation as a core part of their social identity. This 'socialisation' of pseudolaw arguments saw them become not only more standardised but significantly more common, as their use became part of the way members performed allegiance to the respective group.

---

[124] Mark Pitcavage, 'Foreword' in Harry Hobbs, Stephen Young and Joe McIntyre (eds), *Pseudolaw and Sovereign Citizens* (Hart Publishing, 2025) i, vi.

[125] Stephen Young, Harry Hobbs and Rachel Goldwasser, 'The Rise of Sovereign Citizen Pseudolaw in the United States of America' in Harry Hobbs, Stephen Young and Joe McIntyre (eds), *Pseudolaw and Sovereign Citizens* (Hart Publishing, 2025) 95.

[126] Harry Hobbs, Stephen Young and Joe McIntyre, 'The Internationalisation of Pseudolaw: The Growth of Sovereign Citizen Arguments in Australia and Aotearoa New Zealand' (2024) 47(1) *University of New South Wales Law Journal* 309; Glen Cash, 'A Kind of Magic: The Origins and Culture of "Pseudolaw"' (Paper delivered to the Queensland Magistrates' State Conference, Brisbane, 26 May 2022); Stephen Young, Harry Hobbs and Joe McIntyre, 'The Growth of Pseudolaw and Sovereign Citizens in Aotearoa New Zealand' (2023) 1 *New Zealand Law Journal* 6; *Kelly v Fiander* [2023] WASC 187 (1 June 2023); *Rossiter v Adelaide City Council* [2020] SASC 61 (23 April 2020); *Deputy Commissioner of Taxation v Casley* [2017] WASC 161, [15] (Le Miere J).

[127] Mark Pitcavage, 'Foreword' in Harry Hobbs, Stephen Young and Joe McIntyre (eds), *Pseudolaw and Sovereign Citizens* (Hart Publishing, 2025) i, vi.

[128] Frank Chung, "Objectively Nonsense': Sovereign Citizen 'Pseudo-Law' Arguments Clogging Up Australia's Courts', *news.com.au* (online at 16 February 2025) <https://www.news.com.au/finance/economy/australian-economy/objectively-nonsense-sovereign-citizen-pseudolaw-arguments-clogging-up-australias-courts/news-story/79f0627f255bc3c73c9e246e302a5e74>.

[129] Joe McIntyre et al, *The Rise of Pseudolaw in South Australia: An Empirical Analysis of the Emergence and Impact of Pseudolaw on South Australia's Courts* (Final Report, University of South Australia, September 2024).

[130] Mark Pitcavage, 'Foreword' in Harry Hobbs, Stephen Young and Joe McIntyre (eds), *Pseudolaw and Sovereign Citizens* (Hart Publishing, 2025) i, vi.

Perhaps the most significant and influential of these movements has been the 'Sovereign Citizen' movement, which emerged in the United States in the 1990s and remains the largest and most familiar pseudolaw sect. The contemporary Sovereign Citizen movement was itself the progeny of a range of earlier social movements in the United States, including the loosely organised far-right *Posse Comitatus* social movement (founded by William Potter Gale in the early 1970s[131])[132] and the subsequent Common Law Movement (that began to coalesce in the 1980s[133] and promoted a 'radical version of social contract theory'[134]). While the decentralised nature and lack of organisational hierarchy make it impossible to accurately state the precise number of adherents, reports suggest there are now upwards of 500,000 sovereign citizens in the United States.[135]

In the last 20 years, pseudolaw ideology has spread across the globe, with adherents utilising sovereign citizen arguments appearing, for example, in Australia,[136] New Zealand,[137] Canada,[138] the UK,[139] the Netherlands,[140] and Germany.[141] As Pitcavage notes, the 'development of the sovereign citizen movement has allowed numerous arguments and tactics based on pseudolaw to become global in nature and to persist, sometimes for generations.'[142] It was the coalescence of individuals with pseudolegal beliefs forming into larger collective movements that has been at the heart of the global spread of pseudolaw,[143] and the sovereign citizen movement has been at the forefront of this process.

---

[131] Francis Sullivan, 'The Usurping Octopus of Jurisdictional/Authority: The Legal Theories of the Sovereign Citizen Movement' (1999) 4 *Wisconsin Law Review* 785.

[132] It is worth noting that the *Posse Comitatus* itself evolved from a number of earlier movements, most notable the Christian Identity movement (see M Barkun, *Religion and the Racist Right: The Origins of the Christian Identity Movement* (University of North Carolina Press, 1997)). William Potter Gale adapted the pseudo-religious and far-right political ideology of the Christian Identity movement to develop a quasi-legalistic ideology that focused on protecting the rights of individuals: See: Stephen Young, Harry Hobbs and Rachel Goldwasser, 'The Rise of Sovereign Citizen Pseudolaw in the United States of America' in Harry Hobbs, Stephen Young and Joe McIntyre (ed), *Pseudolaw and Sovereign Citizens* (Hart Publishing, 2025) 95.

[133] Harry Hobbs, Stephen Young and Joe McIntyre, 'The Internationalisation of Pseudolaw: The Growth of Sovereign Citizen Arguments in Australia and Aotearoa New Zealand' (2024) 47(1) *University of New South Wales Law Journal* 309, 317.

[134] Daniel Lessard Levin and Michael W Mitchell, 'A Law unto Themselves: The Ideology of the Common Law Court Movement' (1999) 44(1) *South Dakota Law Review* 9, 12.

[135] Kevin Krause, 'What are Sovereign Citizens and What do they Believe?', *The Dallas Morning News* (online at 6 September 2022) <https://www.dallasnews.com/news/politics/2022/09/06/what-is-a-sovereign-citizen-and-what-do-they-believe/>. See Harry Hobbs, Stephen Young and Joe McIntyre, 'The Internationalisation of Pseudolaw: The Growth of Sovereign Citizen Arguments in Australia and Aotearoa New Zealand' (2024) 47(1) *University of New South Wales Law Journal* 309, 318.

[136] Glen Cash, 'A Kind of Magic: Pseudolaw in Australia' in Harry Hobbs, Stephen Young and Joe McIntyre (eds), *Pseudolaw and Sovereign Citizens* (Hart Publishing, 2025) 149; Marilyn McMahon, 'Asserting Sovereignty: An Empirical Analysis of Sovereign Citizen Litigation in Australian Courts' in Harry Hobbs, Stephen Young and Joe McIntyre (eds), *Pseudolaw and Sovereign Citizens* (Hart Publishing, 2025) 175;Harry Hobbs, Stephen Young and Joe McIntyre, 'The Internationalisation of Pseudolaw: The Growth of Sovereign Citizen Arguments in Australia and Aotearoa New Zealand' (2024) 47(1) *University of New South Wales Law Journal* 309.

[137] Harry Hobbs, Stephen Young and Joe McIntyre, 'The Internationalisation of Pseudolaw: The Growth of Sovereign Citizen Arguments in Australia and Aotearoa New Zealand' (2024) 47(1) *University of New South Wales Law Journal* 309; Stephen Young, Harry Hobbs and Joe McIntyre, 'The Growth of Pseudolaw and Sovereign Citizens in Aotearoa New Zealand Courts' (2023) *New Zealand Law Journal* 1.

[138] Donald J Netolitzky,'A Rebellion of Furious Paper: Pseudolaw as a Revolutionary System'(Conference Paper, Sovereign Citizens in Canada Symposium, Centre d'expertise et de formation sur les intégris mes religieux et la radicalisation, 3 May 2018) 1 https://doi.org/10.2139/ssrn.3177484.

However, it is no longer always appropriate to describe these movements as 'sovereign citizens', as the ideology has morphed in many of these instances, and new groupings and social movements that share common legalistic approaches have emerged. For example, other social movements that utilise pseudolegal arguments (often drawing from tropes used by sovereign citizens) include:

- *American State Nationals* (US)[144],
- *Moorish Sovereign Movement* (US)[145]
- The *Freemen-on-the-Land* movement (Canada & UK)[146]
- *Detaxers* (US and Canada)[147]
- the *Magna Carta Lawful Rebellion* (Canada)[148]
- The *Reichsbürger*[149] and '*Sovereignism*' movements (Germany),[150]

---

[139] See, eg, David Griffin and Dana Roemling, 'Signs of Legal and Pseudolegal Authority: A Corpus-Based Comparison of Contemporary Courtroom Filings' (2024) *International Journal for the Semiotics of Law* <https://doi.org/10.1007/s11196-024-10183-7>; David Griffin, '"I Hereby and Herein Claim Liberties": Identity and Power in Sovereign Citizen Pseudolegal Courtroom Filings' (2023) 6 *International Journal of Coercion, Abuse, and Manipulation* 1 <https://doi.org/10.54208/1000/0006/007>.

[140] Luuk de Boer, 'Limit Cases: Sovereign Citizens and a Jurisprudence of Consequences' in M Hertogh and P de Winter (eds) *Empirical Perspectives on the Effects of Law: Towards Jurisprudence and Consequences* <https://figshare.com/articles/journal_contribution/de_Boer_L_O_forthcoming_b_Limit_Cases_Sovereign_Citizens_and_a_Jurisprudence_of_Consequences_In_M_Hertogh_and_P_de_Winter_eds_i_Empirical_Perspectives_on_the_Effects_of_Law_Towards_Jurisprudence_of_Consequences_i_/28275701?file=51914321>.

[141] Anna Lobbert, "Germanite is a Rare Mineral': Sovereignism in Germany' in Harry Hobbs, Stephen Young and Joe McIntyre (ed), *Pseudolaw and Sovereign Citizens* (Hart Publishing, 2025) 203.

[142] Mark Pitcavage, 'Foreword' in in Harry Hobbs, Stephen Young and Joe McIntyre (eds), *Pseudolaw and Sovereign Citizens* (Hart Publishing, 2025) vii.

[143] P usefully explains this process in the following terms: 'Unlike an isolated individual who creates and employs a pseudo-legal concept for their own use, movements can create larger and more complex arguments and use them to recruit thousands of people and teach them harmful tactics based on pseudolaw': Mark Pitcavage, 'Foreword' in in Harry Hobbs, Stephen Young and Joe McIntyre (eds), *Pseudolaw and Sovereign Citizens* (Hart Publishing, 2025) vi.

[144] Christine M Sarteschi, 'American State Nationals: The Next Iteration of the Sovereign Citizen Movement' in Harry Hobbs, Stephen Young and Joe McIntyre (ed), *Pseudolaw and Sovereign Citizens* (Hart Publishing, 2025) 227.

[145] 'Moorish Sovereign Citizens', *SPL Center* (Web Page) <https://www.splcenter.org/resources/extremist-files/moorish-sovereign-citizens/>.

[146] Donald J. Netolitzky, 'The Sun Only Shines on YouTube: The Marginal Presence of Pseudolaw in Canada' in Harry Hobbs, Stephen Young and Joe McIntyre (ed), *Pseudolaw and Sovereign Citizens* (Hart Publishing, 2025) 121, 126-9; *Mead v Mead* [2013] 3 WWR 419, [40].

[147] Donald J. Netolitzky, 'The Sun Only Shines on YouTube: The Marginal Presence of Pseudolaw in Canada' in Harry Hobbs, Stephen Young and Joe McIntyre (ed), *Pseudolaw and Sovereign Citizens* (Hart Publishing, 2025) 121, 129-30.

[148] Ibid 134-5.

[149] See Eric Campbell, 'How Germany's '*Reichsbürger'* Sovereign Citizens Movement Became a Threat to the State' *ABC Foreign Correspondent* (online at 31 August 2023 ) <www.abc.net.au/news/2023-08-31/germany-citizens-of-the-reich-foreign-correspondent/102789818>.

[150] Anna Lobbert, "'Germanite is a Rare Mineral"': Sovereignism in Germany' in Harry Hobbs, Stephen Young and Joe McIntyre (ed), *Pseudolaw and Sovereign Citizens* (Hart Publishing, 2025) 203.

- The '*Kingdom of Gaia*' in (Italy),[151]
- The *Union of Slavic Forces* (Russia)[152]
- *Micronation* proponents (global)[153]

This list is in no way exhaustive, not least because new movements emerge and recede quite rapidly,[154] and because most such movements are inherently fragmented and decentralised, and lack homogeneous ideologies. For this reason, it is an 'oversimplification that discourages further consideration'[155] to utilise 'sovereign citizens' an umbrella term for these movements, and I adopt McRobert's view that 'pseudolaw' is the 'more useful and precise term'[156] to refer to the broader social phenomenon that includes the many sects and manifestations described above.

Understood this way, contemporary pseudolaw is more of an emergent social phenomenon than a monolithic movement set of groups. It is often driven by charismatic local 'gurus' who market ideas to their followers (often for hefty fees),[157] or by subscription-service websites[158] from which adherents can pick and choose arguments. As a result, many users of pseudolaw do not neatly align with a specific group or, indeed may not self-identify as being a member of any broader movement.

## 2. The Nature of Pseudolaw

Despite the variety of manifestations outlined above, there remains a common core to these movements, with adherents advancing novel, indeed fanciful, arguments which they believe (a) represent the 'true' content of the law and (b) allow them to legally avoid undesirable obligations imposed by law. These arguments often refer to legitimate legal sources (often including the *United States Constitution*, Magna Carta, the English *Bill of Rights 1688*, or the United Nations *Universal Declaration of Human Rights*)[159] but use them in novel and unpredictable ways.

As a consequence, one way of understanding pseudolaw is that adherents deploy 'a collection of legal sounding but false rules that purport to be law',[160] but do so in the belief that they are *actually engaged in genuine legal argumentation*. As Netolitzky notes, pseudolaw 'superficially appears to be law … and … uses legal or legal-sounding language but is otherwise spurious'.[161] As I have argued previously, pseudolaw can be identified by three key components:[162]

---

[151] Leonardo Bianchi, 'Dentro l'assurdo mondo dei "sovranisti individuali" italiani', *Vice* (online at 3 June 2017) <https://www.vice.com/it/article/jpn8py/dentro-lassurdo-mondo-dei-sovranisti-individuali-italiani>.

[152] 'Union of Slavic Forces of Russia', *Wikipedia* (Web Page) < https://en.wikipedia.org/wiki/Union_of_Slavic_Forces_of_Russia>.

[153] Harry Hobbs and George Williams, *Micronations and the Search for Sovereignty* (Cambridge University Press, 2021).

[154] Donald J. Netolitzky, 'The Sun Only Shines on YouTube: The Marginal Presence of Pseudolaw in Canada' in Harry Hobbs, Stephen Young and Joe McIntyre (ed), *Pseudolaw and Sovereign Citizens* (Hart Publishing, 2025) 121.

[155] Colin McRoberts, 'Tinfoil Hats and Powdered Wigs: Thoughts on Pseudolaw' (2019) 58 *Washburn Law Journal* 637, 638.

[156] Colin McRoberts, 'Tinfoil Hats and Powdered Wigs: Thoughts on Pseudolaw' (2019) 58 *Washburn Law Journal* 637, 638.

[157] For an illustration of how gurus operate in this context, see *Meads v Meads* [2012] ABQB 571, [19]–[37].

[158] David Heilpern, 'Traffic Matters and Pseudolaw: The Big Shakedown', in Harry Hobbs, Stephen Young and Joe McIntyre (ed), *Pseudolaw and Sovereign Citizens* (Hart Publishing, 2025) 249.

[159] Harry Hobbs, Stephen Young and Joe McIntyre, 'The Internationalisation of Pseudolaw: The Growth of Sovereign Citizen Arguments in Australia and Aotearoa New Zealand' (2024) 47(1) *University of New South Wales Law Journal* 309, 312.

[160] Donald Netolitzky, 'A Rebellion of Furious Paper: Pseudolaw as a Revolutionary System' (Paper delivered to the Centre d'expertise et de formation sur les intégrismes religieux et la radicalisation (CEFIR) symposium: 'Sovereign Citizens in Canada', Montreal, 3 May 2018).

(1) *A Co-opted Legal Form*: it borrows legal terminology and forms of reasoning to *appear* like conventional legal argumentation,[163] creating the illusion of legal legitimacy;

(2) *Use of Contra-Narratives/Alternative Legal Universe*: pseudolaw creates an 'alternate legal universe'[164] that draws from, yet distorts, traditional legal sources to develop its own set of rules; and

(3) *Internalised Belief in the 'True Law'*: it engenders in adherents a genuine belief that they are upholding the 'true law' and can use this knowledge to lawfully achieve their goals.

While adherents are often motivated by an underlying distrust of government and authority, pseudolaw is not anarchistic. Indeed, many believers possess an 'almost endearing commitment to legality and the rule of law.'[165] Instead, there is a belief that by correctly applying their version of legal reasoning in a precise manner, they can achieve miraculous results – avoiding taxes, debts, fines, and other legal liabilities *without breaking the law*. For this reason, pseudolaw is sometimes described as akin to a form of 'magic'.[166] Griffith and Roemling usefully describe this dynamic in the following terms:

> Sovereign Citizens believe that by harnessing the methods purportedly used by those secretive omnipotent individuals, they can force the government and its representatives to do (or not do) anything they desire, including give them access to secret government funds or dismiss criminal charges against them[167]

---

This 'magical thinking' helps to make sense of the rise of pseudolaw. There is increasing recognition[168] that a sense of 'alienation'[169] and 'legal powerlessness'[170] underlies the emergence of pseudolaw. Pseudolaw flourishes where a lack of 'legal literacy'[171] (that is, the ability to meaningfully utilise and understand law and legal processes) collides with the need of an individual to engage with legal institutions.[172] In such situations, an individual who has been alienated from the processes of the law (through lack of legal literacy and meaningful access) is forced to participate in legal processes (to avoid taxes, mortgage repayments, fines, rates or family law obligations). Pseudolaw offers an alluring solution: it promises to 'magically transform the absence of legal power into the presence of legal power.'[173]

Of course, pseudolaw arguments do not work.[174] They are *not* law and are not effective in avoiding legal obligations. Yet the arguments, forms and rituals of pseudolaw often *look* like law to the legal outsider. As Griffith and Roemling note, 'as long as a given pseudolegal text "sounds" sufficiently legal, a layperson may well interpret it as such'.[175] Pseudolaw operates to create an illusion of law, and it flourishes where alienated individuals desperately want to believe in that illusion.

## 3. The Forms of Pseudolaw

As outlined above, pseudolaw uses the forms of legal reasoning, but exists in its own 'legal universe' where it deploys entirely separate substantive arguments. To gain a better understanding of the way in which pseudolaw operates, it is necessary to have some understanding of these various forms of pseudolegal argumentation.

However, pseudolaw adherents do not just mimic the form of legal *arguments*, but the forms of juridical/legalistic behaviour and ritual. Just as traditional legal processes utilise extensive formalistic seeming documents and processes, pseudolegal adherents often deploy a range of ritualistic and theatrical behaviours which they see as having legal significance.

---

[168] Joe McIntyre et al, *The Rise of Pseudolaw in South Australia: An Empirical Analysis of the Emergence and Impact of Pseudolaw on South Australia's Courts* (Final Report, University of South Australia, September 2024); Harry Hobbs, Stephen Young and Joe McIntyre, 'The Internationalisation of Pseudolaw: The Growth of Sovereign Citizen Arguments in Australia and Aotearoa New Zealand' (2024) 47(1) *University of New South Wales Law Journal* 309.

[169] Luuk de Boer, 'Limit Cases: Sovereign Citizens and a Jurisprudence of Consequences' in M Hertogh and P de Winter (eds) *Empirical Perspectives on the Effects of Law: Towards Jurisprudence and Consequences* <https://figshare.com/articles/journal_contribution/de_Boer_L_O_forthcoming_b_Limit_Cases_Sovereign_Cit izens_and_a_Jurisprudence_of_Consequences_In_M_Hertogh_and_P_de_Winter_eds_i_Empirical_Perspectiv es_on_the_Effects_of_Law_Towards_Jurisprudence_of_Consequences_i_/28275701?file=51914321>.

[170] James Gibson and Gregory Caldeira, 'The Legal Cultures of Europe' (1996) 30(1) *Law & Society Review* 55, 65.

[171] Joe McIntyre and Jacqueline Charles, *Submission No 92 to Inquiry into Civics Education, Engagement, and Participation in Australia,* Joint Standing Committee on Electoral Matters, 29 May 2024.

[172] Joe McIntyre et al, *The Rise of Pseudolaw in South Australia: An Empirical Analysis of the Emergence and Impact of Pseudolaw on South Australia's Courts* (Final Report, University of South Australia, September 2024) <https://papers.ssrn.com/sol3/papers.cfm?abstract_id=4996319>.

[173] Luuk de Boer, 'Limit Cases: Sovereign Citizens and a Jurisprudence of Consequences' in M Hertogh and P de Winter (eds) *Empirical Perspectives on the Effects of Law: Towards Jurisprudence and Consequences* https://figshare.com/articles/journal_contribution/de_Boer_L_O_forthcoming_b_Limit_Cases_Sovereign_Citiz ens_and_a_Jurisprudence_of_Consequences_In_M_Hertogh_and_P_de_Winter_eds_i_Empirical_Perspectives _on_the_Effects_of_Law_Towards_Jurisprudence_of_Consequences_i_/28275701?file=51914321> citing Christoph Schönberger and Sophie Schönberger, *Die Reichsbürger: Ermächtigungsversuche einer gespenstischen Bewegung* (Beck, 2023); Luuk de Boer, 'Soevereinen en autonomen in recht en rechtspraak' (2024) 254 *Nederlands Juristenblad* 300.

[174] Harry Hobbs, Stephen Young and Joe McIntyre, 'The Internationalisation of Pseudolaw: The Growth of Sovereign Citizen Arguments in Australia and Aotearoa New Zealand' (2024) 47(1) *University of New South Wales Law Journal* 309.

[175] David Griffin and Dana Roemling, 'Signs of Legal and Pseudolegal Authority: A Corpus-Based Comparison of Contemporary Courtroom Filings' (2024) *International Journal for the Semiotics of Law* [2.3].

One example of this type of behaviour is the filing of documents that appear – on their face – to reflect the forms of genuine legal documents. Netolitsky illustrates such behaviour in the form of an 'affidavit' filed by an adherent in one case:

**Figure 4: Illustrative Pseudolaw Document:** *Yankson Moorish Law 'Affidavit'*[176]

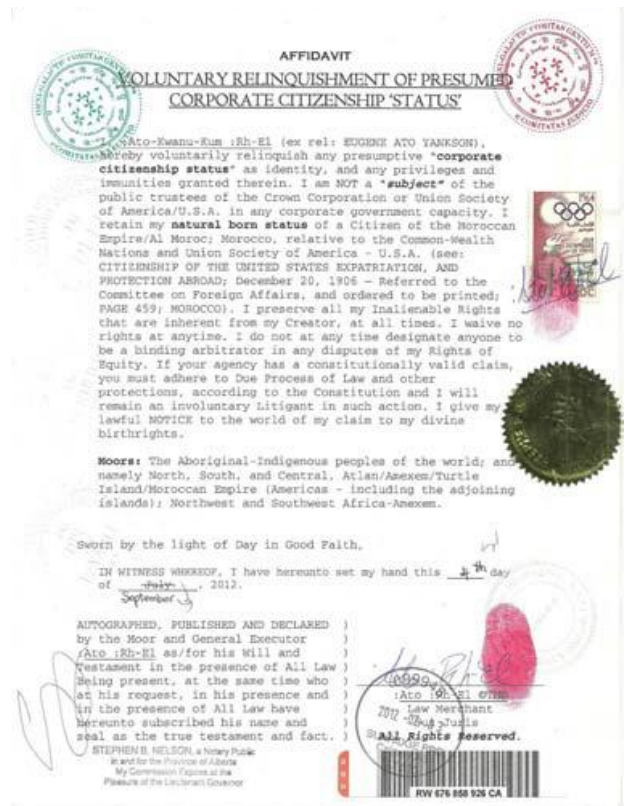

Such a document appears, at first glance, to be a standard formal legal document. To the non-legally trained, it would appear formal and legally significant. It is only on closer inspection that some of the bizarre pseudolaw ritualistic behaviours are revealed, including the affixed postage stamp and red fingerprint.

In the following section, I will briefly outline some of the ways in which pseudolaw manifests, beginning with an overview of common pseudolaw arguments, before outlining some of the significant pseudolaw rituals and common behaviours.

### (a) Pseudolegal Argumentation

At first glance, pseudolaw often seems nonsensical, with courts frequently describing it as 'obvious nonsense',[177] 'pseudo-legal gibberish',[178] or 'gobbledygook'.[179] As I have previously stated, when 'the legally trained encounter pseudolegal arguments, the first instinct is often to dismiss it as hallucination that warrants no further analysis.'[180] Yet, underlying this gibberish is are discernible structures and common argumentative patterns that can be understood.

---

[176] Donald J Netolitzky, 'A Rebellion of Furious Paper: Pseudolaw as a Revolutionary System' (Conference Paper, Sovereign Citizens in Canada Symposium, Centre d'expertise et de formation sur les intégris mes religieux et la radicalisation, 3 May 2018) 1 <https://doi.org/10.2139/ssrn.3177484>.

[177] *Bradley v The Crown* [2020] QCA 252 (13 November 2020).

[178] *Deputy Commissioner of Taxation v Casley* [2017] WASC 161, [15] ('*Casley*').

[179] *National Australia Bank Ltd v Norman* [2012] VSC 14, [4].

One of the earliest and most influential efforts to provide a systematic overview of the various pseudolaw arguments was provided by Rooke ACJ of the Alberta Court of Queen's Bench in the case of *Meads v Meads*.[181] In a contentious family law dispute, one of the parties had submitted a collection of pseudolegal tropes and arguments. In a 160-page, 736-paragraph decision, Rooke ACJ provided a 'magisterial'[182] overview and refutation of pseudolaw as it existed in Canada. Rooke ACJ recognised that despite the various labels used by respective movements, there were common themes and arguments, and developed the term 'Organized Pseudolegal Commercial Argument' ('OPCA') to refer to the broader phenomenon.[183] He then outlined what he saw as the five main forms of OPCA type arguments, namely: (a) The 'Litigant is Not Subject to Court Authority' Argument;[184] (b) The 'Obligation Requires Agreement' Argument;[185] (c) The 'Double/Split Persons' Argument;[186] (d) The 'Unilateral Agreements' Argument;[187] and (e) 'Money for Nothing' Schemes/Arguments.[188]

Building on this work, Donald Netolitzky[189] developed a typology he described as the 'pseudolaw memeplex' which has six core concepts: (1) Everything is a contract; (2) Silence means acceptance or agreement; (3) Legal action requires that there be an 'injured party'; (4) Government authority is defective or at least limited; (5) The 'Strawman' duality; and (6) Financial and banking conspiracy theories.[190] For Netolitzky, these core ideas have possessed an almost pathogen-like quality as they have subsequently been adapted and applied by pseudolaw adherents across the globe.

An example of how these ideas have spread is provided by works I co-authored in 2023-24, which undertook the doctrinal analysis of the forms of pseudolegal argumentation as they appear in the reported judgments of Australia and Aotearoa New Zealand.[191] In those works, we initially identified three broad categories of arguments:

---

[180] Joe McIntyre, Frankie Bray and Madeleine Perrett, 'The Quantitative Analysis of the Rise of Pseudolaw in South Australia' (forthcoming, 2025)

[181] *Meads v Meads* [2012] ABQB 571.

[182] Harry Hobbs, Stephen Young, and Joe McIntyre, 'The Internationalisation of Pseudolaw: The Growth of Sovereign Citizen Arguments in Australia and Aotearoa New Zealand' (2024) 47(1) *UNSW Law Journal* 309, 316.

[183] *Meads v Meads* [2012] ABQB 571, [1].

[184] Ibid [62]–[87].

[185] Ibid [87]–[93].

[186] Ibid [93]-[101]. This is now understood as the 'strawman' argument: See generally   Joe McIntyre, Harry Hobbs and Stephen Young, 'The Strawmen Trap: Non-appearance and the Pitfalls of Pseudolaw,' (2025) 99 *Australian Law Journal* 1

[187] Ibid [101]–[120].

[188] Ibid [120]–[124].

[189] It is worth noting that Netolitzky was a judicial administrator in the Alberta Court of Queen's Bench and worked on Mead. He has subsequently become one of the preeminent pseudolaw scholars globally.

[190] Donald J Netolitzky, 'A Rebellion of Furious Paper: Pseudolaw as a Revolutionary System'(Conference Paper, Sovereign Citizens in Canada Symposium, Centre d'expertise et de formation sur les intégris mes religieux et la radicalisation, 3 May 2018) 1 <https://doi.org/10.2139/ssrn.3177484>.

[191] Harry Hobbs, Stephen Young, and Joe McIntyre, 'The Internationalisation of Pseudolaw: The Growth of Sovereign Citizen Arguments in Australia and Aotearoa New Zealand' (2024) 47(1) *UNSW Law Journal* 309; Stephen Young, Harry Hobbs and Joe McIntyre, 'The Growth of Pseudolaw and Sovereign Citizens in Aotearoa New Zealand' (2023) 1 *New Zealand Law Journal* 6; Joe McIntyre et al, *The Rise of Pseudolaw in South Australia: An Empirical Analysis of the Emergence and Impact of Pseudolaw on South Australia's Courts* (Final Report, University of South Australia, September 2024).

(1) *The Strawman Argument:* the law does not apply because it applies only to 'artificial' persons who possess a separate legal personality – the strawman duality;

(2) *Absence of Individual Consent*: government authority is illegitimate in the absence of individual consent, and they did not consent to the law operating upon them – everything is a contract; and/or

(3) *State Law is Defective: the* law was invalidly enacted and is of no legal effect – state authority is defective or limited.[192]

In later research, we identified a number of additional tropes and forms of argument, including private prosecutions,[193] and the use of a range of pseudolegal traffic law arguments.[194]

While it is not necessary in this paper to provide an overview of how all these various pseudolaw arguments operate, it is useful to illustrate how these operate by reference to one of these common argumentative tropes. The following section briefly outlines one of the most notorious pseudolaw arguments: the strawman theory.[195]

The 'strawman' theory was created and popularised by North Dakota farmer and sovereign citizen, Roger Elvick in the late 1990s.[196] The argument purports to empower adherents to avoid legal obligations through the performance of certain ritualistic behaviours. The argument has three key components. The first is the idea that an individual has two personas, 'one of himself as a real flesh and blood human being and the other, a separate legal personality who is the straw man'.[197] As Vandogen J noted in *Kelly v Fiander*, the strawman theory 'is based on the fundamentally misguided notion that there exists a physical human being and, at the same time, a separate non-physical person (a "doppelganger")'.[198] The second component is that different legal rights, responsibilities, and obligations adhere to the strawman/doppelganger as opposed to the natural person.[199] Adherents will often use specific ritualistic forms, such as writing names in all capitalisation, to differentiate between the two personas.[200] The third aspect is that through certain

---

[192] Harry Hobbs, Stephen Young, and Joe McIntyre, 'The Internationalisation of Pseudolaw: The Growth of Sovereign Citizen Arguments in Australia and Aotearoa New Zealand' (2024) 47(1) *UNSW Law Journal* 309, 324.

[193] Joe McIntyre et al, *The Rise of Pseudolaw in South Australia: An Empirical Analysis of the Emergence and Impact of Pseudolaw on South Australia's Courts* (Final Report, University of South Australia, September 2024).

[194] Joe McIntyre et al, *The Rise of Pseudolaw in South Australia: An Empirical Analysis of the Emergence and Impact of Pseudolaw on South Australia's Courts* (Final Report, University of South Australia, September 2024); David Heilpern, 'Traffic Matters and Pseudolaw: The Big Shakedown', in Harry Hobbs, Stephen Young and Joe McIntyre (ed), *Pseudolaw and Sovereign Citizens* (Hart Publishing, 2025) 249.

[195] For an overview of this argument, and how it can create problems for courts, see Joe McIntyre, Harry Hobbs and Stephen Young, 'The Strawmen Trap: Non-appearance and the Pitfalls of Pseudolaw,' (2025) 99 *Australian Law Journal* 1

[196] Joe McIntyre, Harry Hobbs and Stephen Young, 'The Strawmen Trap: Non-appearance and the Pitfalls of Pseudolaw,' (2025) 99 *Australian Law Journal* 1, 5

[197] *Deputy Commissioner of Taxation v Casley* [2017] WASC 161, [15]

[198] *Kelly v Fiander* [2023] WASC 187 (1 June 2023) [11]. In *Meads v Meads*, Rooke ACJ observed: 'This confusing concept is expressed in many different ways. The 'physical person' is one aspect of the duality, the other is a non-corporeal aspect that has many names, such as a 'strawman', a 'corporation', a 'corporate entity', a 'corporate fiction', a 'dead corporation', a 'dead person', an 'estate', a 'legal person', a 'legal fiction', an 'artificial entity', a 'procedural phantom', 'abandoned paper work', a 'slave name' or 'slave person', or a 'juristic person.'' *Meads v Meads* [2012] ABQB 571 [417].

[199] Joe McIntyre, Harry Hobbs and Stephen Young, 'The Strawmen Trap: Non-appearance and the Pitfalls of Pseudolaw,' (2025) 99 *Australian Law Journal* 1, 7

[200] A classic example comes from *R v Sweet*: 'The essence of the applicant's argument is that he possesses two distinct personas. One the 'real live flesh and blood man' and the other a 'straw man' or 'dummy corporation'. The former is designated in the applicant's material as 'Kym-Anthony:' and the latter as KYM ANTHONY

ritualistic actions, these two personas can be severed, so the legal obligations stay with the strawman/doppelganger and the initial subjugation of the natural person is negated. As Vandogen J noted in *Kelly v Fiander*:

> A critical component of this strawman theory is the idea that government authority over the physical person can be negated by removing the doppelganger. In very simple terms, this is said to be achieved by revoking or denying the legitimacy of the contract. This then has the effect of removing any government authority over the physical person.[201]

There is a wide range of pseudolegal rituals that are deployed by adherents in an attempt to affect this severance between different personas. These commonly include trying to get the court to recognise their assertion of natural sovereignty, for example, by acknowledging that they appear as a 'flesh and blood man' or 'individual person',[202] or as the 'executor' of the artificial strawman.[203] The pseudolegal belief is that once this separation is acknowledged, the 'natural person' is freed from the preceding legal obligations and thus no longer has to pay taxes, debts, etc.

Adherents attempt to support the various arguments by reference to cases and statutes, presenting arguments in ways that utilise the form of law. Yet, ultimately, these arguments do not use the intellectual architecture and modes of reasoning[204] that are the central determinants of legal argumentation. Pseudolaw arguments are presented confidently, yet they are not legally effective.

### *(b) Rituals and Behaviour*

In many respects, though, pseudolaw is best understood through the behaviours and attitudes of its followers, rather than simply through the style and content of arguments used. These behaviours include a conspiratorial worldview, a constant questioning of authority, the proliferation of irrelevant and voluminous filings, and the use of scripted, often illogical arguments. Moreover, a degree of theatricality, indeed of ritualistic behaviour, has been recognised as one of the defining features of pseudolaw,[205] and this theatrical nature of pseudolaw is becoming better understood in the literature.[206] For example, in his influential paper 'A Kind of Magic', Cash observes:

> Ritual and ceremony have long been at the heart of pseudolaw ideology. Documents are marked with signals and signs. Written submissions bear the appearance of incantations. Statutes are parsed to discover hidden meaning and codes. It is unsurprising then that pseudolaw has been likened to magic.[207]

---

SWEET. According to the applicant's argument, the real person is not subject to the laws of Queensland, and the charges should be dismissed.' *R v Sweet* [2021] QDC 216, [2] (references removed).

[201] *Kelly v Fiander* [2023] WASC 187 [13].

[202] *Deputy Commissioner of Taxation v Cutts (No.4)* [2019] FCCA 2866 (10 October 2019) [127] ('*Cutts*').

[203] *Kelly v Fiander* [2023] WASC 187 [17].

[204] See Joe McIntyre, *The Judicial Function: Fundamental Principles of Contemporary Judging* (Springer, 2019)

[205] Harry Hobbs, Stephen Young, and Joe McIntyre, 'The Internationalisation of Pseudolaw: The Growth of Sovereign Citizen Arguments in Australia and Aotearoa New Zealand' (2024) 47(1) *UNSW Law Journal* 309, 313.

[206] Ibid, 313; Kate Leader, 'Conspiracy! Or, When Bad Things Happen to Good Litigants in Person' (2024) 44(3) *Legal Studies* 1. See also Kate Leader, *Litigants in Person in the Civil Justice System: In Their Own Words* (Hart, 2024).

[207] Glen Cash, 'A Kind of Magic: The Origins and Culture of "Pseudolaw"' (Speech Delivered at the Queensland Magistrates State Conference, Brisbane, 2022) 9.

Cash has subsequently provided, by reference to the filings of a litigant in the case of *R v Sweet*,[208] a useful illustration of the way symbols and rituals of pseudolaw are deployed by adherents.[209] These documents 'weighed down with magical portent'[210] are best appreciated by their reproduction:

**Figure 5: Illustrative Pseudolaw Document Filed in Court[211]**

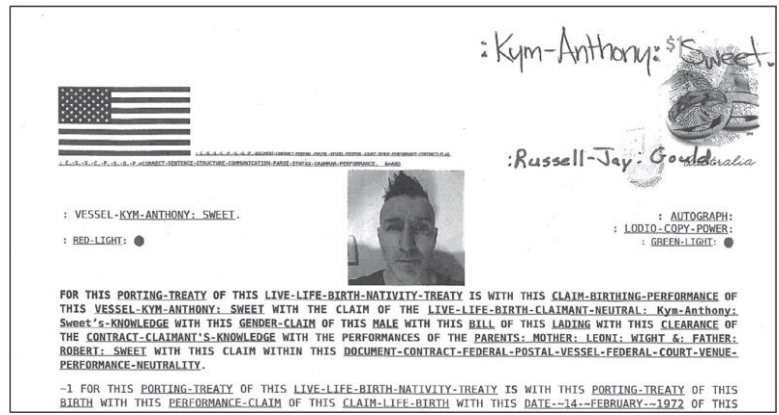

In the figure above, key ritualistic aspects include the use of all capitalisations, the attachment of a postage stamp, the affixation of a fingerprint and a photograph, the inclusion of an image of a flag and the reference to his name in the form 'Kym-Anthony: Sweet'. Some of these same behaviours are evidenced in a 'birth certificate' also filed in that case:

**Figure 6: Illustrative Pseudolaw 'Birth Certificate' Filed in Court[212]**

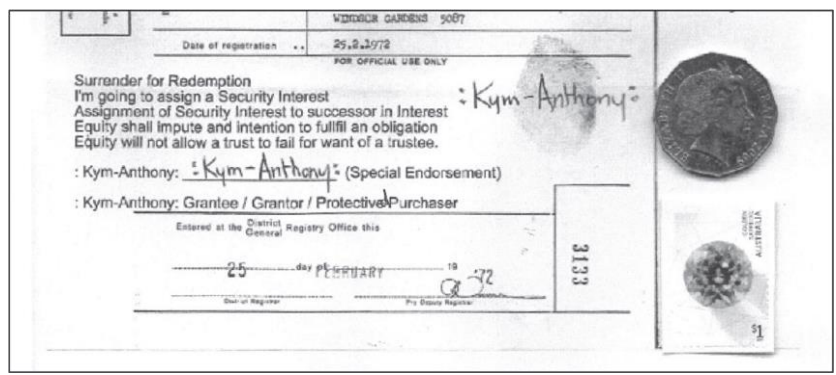

Note that in this document, the litigant attached not only a postage stamp but also a 50-cent coin. In another 'filing', the litigant submitted a £2 coin commemorating the visit of Queen Elizabeth II to Alderney (a Channel Island) in 1989.[213] As Cash notes, these behaviours were all part of the

---

'performance of pseudolaw ritual', with the 'combination of coins, stamps and fingerprints deployed by Sweet in a single case … illustrative of how deeply forms and rituals of this kind have become embedded in the pseudolaw movement.'[214]

Another notable behaviour feature is the reliance on legal instruments that are formally irrelevant to the case at hand – such as references to *Magna Carta*[215] - that are seen to be somehow significant. As one judicial administrator noted:

> …in the sense that it's *referring to obscure, if they even exist, very obscure, ancient pieces of legislation or common law* that likely were articulated, if they were ever articulated, or in a context that has nothing to do with what you are dealing with now. But *it's a set of words that sounds good for the circumstances*.[216]

This approach highlights that arguments were often not particularly targeted or relevant. Indeed, participants noted that pseudolaw litigants seemed to rely on common scripts or templates. As one judicial administrator noted, there seems to be the 'same sort of template that everyone's using'.[217] Notably, it appears that adherents often do not understand the significance of the rituals and behaviours they are deploying, only knowing that they *are* important. As Cash notes, the significance of the (abovementioned) commemorative coin was not made apparent and 'it was never mentioned during argument – *and I suspect even Sweet could not explain its meaning.*'[218] Similarly, other evidence suggests some adherents appear to simply read out a 'script' in litigation without rehearsal or any real understanding of what they were saying (or its implications).[219]

Another common form by which pseudolaw manifested was in the extensive use of paperwork and voluminous filings by adherents in their litigation. In our recent empirical study,[220] One Supreme Court judicial officer remarked in an interview that:

> The *first indicator is always the voluminous filings*. Enormous quantities of documents and pleadings and then trying to work out what the issues are from those pleadings is almost impossible … *Very keen on paperwork, very keen*.[221]

These behaviours are, in many respects, as significant in the distinction of pseudolaw as the substance of the core pseudolegal arguments and ideology. Though the study of the semiotics of

---

pseudolaw is in its infancy,[222] pseudolaw litigants are acting differently in court compared to 'normal' litigants.

However, despite these behaviours deviating significantly from traditional litigious performance and behaviour, there is a sense in which pseudolaw behaviour looks like what you may expect legally significant behaviour to look like *if you didn't understand legal practice*. Legal practice *is* replete with impenetrable language, reference to old and obscure text, reliance on authority, voluminous filings, and often strange rituals. But these behaviours exist without an intellectual infrastructure that makes sense and necessitates them. There is an underlying substantive content and rationale for the behaviour. Pseudolegal behaviour, in contrast, apes the form of legal behaviour but does so without the underlying conceptual framework and substantive rationale. It is as if an actor tried to perform a role *legalistically* without any understanding of why.

Both in argumentative content and in litigious behaviour, pseudolaw is a triumph of form over substance.

## 4. Pseudolaw as a Harmful Mimic

The rise of pseudolaw since 2020 has had significant impacts on courts and other legal institutions all around the world. Our recent empirical analysis suggested that pseudolaw has become a weekly, if not daily, occurrence in many courts,[223] with judicial administrators observing that communications with pseudolaw adherents have become 'effectively continuous'.[224] This is having a profound impact on the good administration of justice; pseudolaw is now seen as one of the most detrimental issues for those working in the court system.[225]

Courts are busy places, and pseudolaw can be particularly disruptive,[226] not least because of the disproportionate impact of each pseudolaw case compared to other cases.[227] The case law involving pseudolaw matters commonly includes statements of judicial concern about the systemic impact of such cases. For example, in *Rossiter v Adelaide City Council*,[228] Livesey J observed that pseudolegal arguments have 'without reservation been rejected as involving both legal nonsense and an unnecessary waste of scarce public and judicial resources.'[229] Similarly, in *Smadu v Stone*,[230] Le Mire J observed that it was a 'waste of judicial resources and an affront to the dignity of [the] court' to be forced to substantively deal with pseudolaw arguments.[231] The burden of

---

dealing with pseudolaw matters is having a profound impact on the courts. As one judicial officer put it:

> It uses up so much court resources because every time they want to file something they end up making the lives of the Registry staff miserable. They send [the] clerk, her email inbox or the Chambers inbox is full of emails from them. *They have changed the whole face of the civil justice system.*[232]

Moreover, some judges have raised concerns that the 'cost' of pseudolaw goes beyond wasted resources to undermine the authority and social licence of courts. For example, in *Yap v Mantic (No 7),* Solomon J observed:

> [the pseudolaw adherent's] behaviour has not been without a cost. That cost involves the court resources that are wasted in hosting repeated hearings when parties fail to appear. There is importantly, a social and reputational cost to the court when litigants are given wide berth to disregard or ignore court orders. There is a cost to other parties, who have the right to expect that court orders will be complied with. … *The growing prevalence of this class of beliefs and associated conduct is a matter of concern for the preservation of the authority of the court and the administration of justice.*[233]

Yet the cost of pseudolaw extends far beyond the courts, with 'broader public disruption of services'[234] for a range of public services, including police, public health, and local governments. This ultimately results in increased enforcement costs and public revenue shortfalls,[235] increasing costs for the general public.[236]

And while it may be difficult to feel sympathy for the adherents themselves, it is also clear that the use of pseudolaw detrimentally impacts those adherents who often 'often end up with a worse outcome.'[237] This issue was recognised by Livesey J when he observed, in *Rossiter,*[238] that:

> It is regrettable that the appellant has advocated the various pseudolegal arguments underpinning this appeal. If he has acted on the advice of others, he is well advised to stop doing so. His decision to defend has resulted in a trivial parking fine escalating to a financial burden exceeding $2,000.[239]

---

[232] Joe McIntyre et al, *The Rise of Pseudolaw in South Australia: An Empirical Analysis of the Emergence and Impact of Pseudolaw on South Australia's Courts* (Final Report, University of South Australia, September 2024) 77.

[233] *Yap v Matic (No 7)* [40]. Similar concerns were expressed in *Nelson v Greenman,* where Justice Gobbo vividly expressed the concerning in the following passage: '…the first defendant's arguments before me concerning the 'Living Man' and 'the People's Court of Terra Australis' were nothing more than carnival of absurdity drawn from a mishmash of delusional arguments. *Whilst it may be tempting to simply dismiss these claims as nonsense, gibberish, gobbledegook or like, in doing so that should not diminish from the serious impact these delusional arguments can have on the authority of the Court.*' *Nelson v Greenman* [2024] VSC 704 (15 November 2024) [70] (Gobbo J).

[234] Joe McIntyre et al, *The Rise of Pseudolaw in South Australia: An Empirical Analysis of the Emergence and Impact of Pseudolaw on South Australia's Courts* (Final Report, University of South Australia, September 2024) Interview 6: Judicial Officer, 576.

[235] Ibid, 580.

[236] Ibid, Interview 6: Judicial Officer, 585 – 590; Interview 3: Judicial Administrator, 621–625.

[237] Ibid, Interview 1: Judicial Officer, 510; Harry Hobbs, Stephen Young, and Joe McIntyre, 'The Internationalisation of Pseudolaw: The Growth of Sovereign Citizen Arguments in Australia and Aotearoa New Zealand' (2024) 47(1) *UNSW Law Journal* 309, 338–40.

[238] *Rossiter v Adelaide City Council* [2020] SASC 61.

[239] *Rossiter v Adelaide City Council* [2020] SASC 61, [52] (Livesey J).

This impact on the litigant becomes relevant when we consider the question of who benefits from the advancement of pseudolaw. While believers may gain psychological benefits from countering feelings of alienation, there is no material advantage – with no pseudolaw argument *ever* succeeding.[240] Yet, as I discuss below, there is money to be made by the gurus and leaders who sell access to believers.

Ultimately, these significant detrimental impacts of pseudolaw highlight the 'alienation and polarisation of individuals in neoliberal states and the difficulties that many face in utilising the legal system and accessing justice'.[241] There are unique characteristics of law that make it particularly vulnerable to form-based mimicry as opposed to other disciplines. As Griffith and Roemling note, law is required to 'simultaneously serve the specific technical needs of a range of professions while theoretically remaining entirely accessible to laypeople,' with the result that it is 'particular is vulnerable to this sort of attempted authoritative appropriation'.[242] Because law *does* use archaic language and often byzantine ritual, it is far too easy for an alienated and excluded layperson to mistake the (pseudolegal) behaviour that mimics the form of law for law itself. As Griffith and Roemling observe, 'as long as a given pseudolegal text "sounds" sufficiently legal, a layperson may well interpret it as such'.[243]

## PART IV: THE TRIUMPH OF FORM OVER SUBSTANCE

In the above sections, I provided an overview of two recently emerging social phenomena, LLMs and pseudolaw, intending to demonstrate that both share common characteristics. In both cases, the public is engaging with an inherently complex and inaccessible forms of technology that produce outputs that *look like what they think the output should look like*.

For LLMs, the mindboggling scale of permutations involved, and the unavoidable opacity of the deep neural networks, mean that the technological function of the system is beyond the comprehension of most people, yet the output appears eerily close to actual language. It becomes far too easy in such a context to treat the pseudo-language outputs of an LLM as language itself, with the inherent elements of intentionality and externality. These LLMs seem to be the intoxicating proposition of the democratisation of all human knowledge and expertise. Similarly, when disaffected and alienated individuals with poor functional legal literacy are confronted with the coercive power of law, the allure of pseudolaw can be compelling.[244] The esoteric and inaccessible law is suddenly reclaimed, and the new rituals and arguments seem to offer an inversion of traditional legal powerlessness. Pseudolaw seems to share all the outward characteristics of law, a seemingly common form of both argumentative style and ritualistic behaviours: it must, therefore, be law! Yet again, what is occurring is a false analogy – pseudolaw is no more law than LLM pseudo-language is genuine communicative language.

---

[240] Harry Hobbs, Stephen Young and Joe McIntyre, 'The Internationalisation of Pseudolaw: The Growth of Sovereign Citizen Arguments in Australia and Aotearoa New Zealand' (2024) 47(1) *University of New South Wales Law Journal* 309.

[241] Maria O'Sullivan, 'Pseudolaw and Legal Fictions: Vaccine Mandate Claims During the COVID-19 Pandemic and Future Implications' in Harry Hobbs, Stephen Young and Joe McIntyre (ed), *Pseudolaw and Sovereign Citizens* (Hart Publishing, 2025) 37, 39; Luuk de Boer, 'Limit Cases: Sovereign Citizens and a Jurisprudence of Consequences' in M Hertogh and P de Winter (eds) Empirical Perspectives on the Effects of Law: Towards Jurisprudence and Consequences <https://figshare.com/articles/journal_contribution/de_Boer_L_O_forthcoming_b_Limit_Cases_Sovereign_Cit izens_and_a_Jurisprudence_of_Consequences_In_M_Hertogh_and_P_de_Winter_eds_i_Empirical_Perspectiv es_on_the_Effects_of_Law_Towards_Jurisprudence_of_Consequences_i_/28275701?file=51914321>.

[242] David Griffin and Dana Roemling, 'Signs of Legal and Pseudolegal Authority: A Corpus-Based Comparison of Contemporary Courtroom Filings' (2024) *International Journal for the Semiotics of Law* [2.3].

[243] David Griffin and Dana Roemling, 'Signs of Legal and Pseudolegal Authority: A Corpus-Based Comparison of Contemporary Courtroom Filings' (2024) *International Journal for the Semiotics of Law* [2.3].

[244] See Luuk de Boer, 'Limit Cases: Sovereign Citizens and a Jurisprudence of Consequences' in M Hertogh and P de Winter (eds) *Empirical Perspectives on the Effects of Law: Towards Jurisprudence and Consequences* <https://figshare.com/articles/journal_contribution/de_Boer_L_O_forthcoming_b_Limit_Cases_Sovereign_Cit

Understood in this way, it should be immediately apparent how both LLMs and pseudolaw share common traits:

- Both succeed in replicating the form of their mimicked output, but not its substance; and
- Both flourish where their users desperately wish to believe in the efficacy of their magical output.

Once these common traits are recognised, it becomes possible to further identify a number of common causes and consequences of both phenomena, and - through the juxtaposition of them – to gain a deeper insight into each phenomenon. Moreover, when we better understand the 'form over substance' nature of both phenomena it becomes easier to identify both when they 'succeed' as well as the risks they pose when they fail.

In the following Part, I will outline four such implications which are common to both phenomena, but which are most clearly exposed when the two phenomena are juxtaposed. These are:

1) **Conceptual Pareidolia:** Firstly, both phenomena draw upon the human tendency to perceive meaningful patterns in unrelated stimuli. In mistaking the surface form for the thing itself, both phenomena succeed in part because of innate cognitive behaviours.

2) **The Mistaken Correlation of Confidence for Competence**: Secondly, both phenomena are distinguished by the particularly confident utilisation of form, and the tendency of the user to mistake that confidence for substantive knowledge. In both cases, this error draws upon human heuristic to treat confidence as a proxy for knowledge (the 'confidence heuristic').

3) **Form-Driven Success:** Thirdly, both phenomena can be highly successful in a certain way, provided the relevant domain focuses on form rather than substance. Explicit recognition of this makes it easier to recognise the appropriate uses and responses to these phenomena

4) **The Use of Magic Thinking:** Finally, both phenomena appear to offer substantive success, and in both cases, there is a perception by users that the use is indeed substantively successful. I argue that in both cases that perceived success emanates from a form of 'magical thinking' by users.

These four implications are present in both phenomena but are significantly more easily recognised and analysed in the juxtaposition between them. In the following section, I will discuss the nature of each of these four implications, before reflecting, in the final part, on what this may mean in the development of appropriate responses.

## 1. Demons and Ghosts: Pareidolia & Apophenia in Pseudolaw & LLMs

---

izens_and_a_Jurisprudence_of_Consequences_In_M_Hertogh_and_P_de_Winter_eds_i_Empirical_Perspectiv es_on_the_Effects_of_Law_Towards_Jurisprudence_of_Consequences_i_/28275701?file=51914321>.

In many ways, the most significant insight to be found in juxtaposing the phenomena of pseudolaw and LLM pseudo-language is that in both cases, a perception of meaningful behaviour is created in the users' mind from inputs that lack such meaning. Essentially, in both cases, users mistake the form for the thing itself, finding meaning and patterns where none exist.

For pseudolaw, this means adherents mistake pseudolegal arguments that share some sources and forms with legal reasoning *with law itself*. Adherents perform legalistic rituals and use legalistic-sounding language. In doing so, they believe that they are doing law. Adherents find apparent patterns in legal sources where in fact there are no such patterns because they lack the fundamental conceptual infrastructure of legal reasoning. They think that so long as form is followed (sources are cited, legalistic language used, ritual performed), then substantive legal norms are generated.

For LLMs that stitch together linguistic forms without any reference to meaning, humans users tend to mistakenly assume that, given the familiar form, the output has meaning that pseudo-language is simply language. In this case, the output has the *form* of language, so users assume it is engaging in communication - *a transfer of meaning*. But as established above, this is an error: LLMs do not produce meaning, only the appearance of it. Yet, users persistently project substantive meaning onto that outputted form. The LLM is treated as a source of knowledge, of substantive meaning, rather than simply a mimic of human language.

### (a) Pareidolia and Apophenia in Cognitive Psychology

This common practice can be analogous to the concept of 'pareidolia' in cognitive psychology. Pareidolia, as it is generally understood, is the tendency to perceive patterns and meaningful interpretation in things that are nebulous and otherwise meaningless.[245] More technically, it involves illusory sensory perception,[246] and occurs when stimuli 'trigger perceptions of non-existent entities' in a manner that reflects 'erroneous matches between internal representations and the sensory inputs'.[247] Most people are familiar with the phenomenon, if not the nomenclature: we experience pareidolia when we see animals in clouds, a wise face in the gnarled trunk of a tree, or a smiling face in the headlights.

Pareidolia is a type of 'apophenia' which is the tendency to 'perceive illusory patterns in random and unconnected events or stimuli,'[248] that is, to 'see a difference or meaning when the given result is attributable to chance'.[249]

The concept was first described by German psychiatrist Klaus Conrad in a paper describing the acute stage of schizophrenia,[250] during which unrelated details seem saturated in connections and meaning.[251] The term was introduced to English by Swiss psychologist Peter Brugger in 2001, who described apophenia as a weakness of human cognition: the "pervasive tendency … to see order in random configurations," an "unmotivated seeing of connections," the experience of "delusion as revelation." [252] Critically, for apophenia, the 'perception of connections or meaning in unrelated events'[253] is seen to bear concrete and specific meaning for the individual. Sandra Hubscher gives the example of an account by Swedish playwright August Strindberg's *Inferno/From an Occult Diary:*

---

[245] A useful definition for public consumption is provided by Thompson, who observes that '[p]areidolia is the phenomenon where people interpret something definite, precise and often meaningful from what is actually random arrangements of stimulus': Jesse Thompson, 'Pareidolia: seeing faces in random, inanimate objects could be survival technique', *ABC News* (online at 17 February 2019) <https://www.abc.net.au/news/2019-02-17/pareidolia-explains-seeing-faces-in-objects/10813426>.

[246] Jiangang Lui et al, 'Seeing Jesus in Toast: Neural and Behavioral Correlates of Face Pareidolia' (2014) 53 *Cortex* 60, 60.

[247] Ibid.

[248] Zack W. Ellerby and Richard J. Tunney, 'The Effects of Heuristics and Apophenia on Probabilistic Choice' (2017) 13(4) *Advances in Cognitive Psychology* 280, 281.

[249] Sandra L Hubscher, 'Apophenia: Definition and Analysis', *Archive.md* (online, 4 November 2007).(

[250] Klaus Conrad, *Die Beginnende Schizophrenie: Versuch einer Gestaltanalyse des Wahns* (Psychiatrie-Verlag, 1959). The translated title of this book is *The Commencing Schizophrenia: Trial on a Gestalt Analysis of Delusion*. In the following year, Conrad wrote '[b]orrowing from ancient Greek, the artificial term 'apophany' describes this process of repetitively and monotonously experiencing abnormal meanings in the entire

> "There on the ground I found two dry twigs, broken off by the wind. They were shaped like the Greek letter for "P" and "y"… [I]t struck me that [they] must be an abbreviation of the name Popoffsky. Now I was sure it was he who was persecuting me, and that the Powers wanted to open my eyes to my danger."[254]

Here, Strindberg seems to be describing an apophany - a false realization of the world's interconnectedness. It is not merely that he perceives a pattern in the twigs but that this pattern possesses a specific and intentional meaning for him, which Hubscher notes is likely indicative of schizophrenia.[255]

There are generally seen to be four main types of apophenia: Pareidolia; Clustering illusion; Confirmation bias; and Gambler's fallacy.[256] Each involves a form of 'false positivity' – a 'Type I error' – of determining that some relationship is meaningful or true when it is not the case. Although apophenia (sometimes described as 'patternicity')[257] may be connected to survival instincts,[258] its central features are that the connections people see are not really there[259] and that the person places unwarranted significance on those connections. While recognising patterns can be a part of natural brain function, apophenia goes beyond this: what was simply a cognitive bias can become a sign of a mental health condition. There is a correlation between psychoticism and

---

surrounding experiential field, eg, being observed, spoken about, the object of eavesdropping, followed by strangers': Klaus Conrad, 'Gestaltanalyse und Daseinsanalytik' (1959) 30 *Nervenarzt* 405 quoted in Aaron L Mishara, 'Klaus Conrad (1905–1961): Delusional Mood, Psychosis, and Beginning Schizophrenia' 36(1) *Schizophrenia Bulletin* 9, 10 <https://pmc.ncbi.nlm.nih.gov/articles/PMC2800156/#ref-list1>.

[251] Katy Waldman, 'It's All Connected', *Slate* (online at 16 September 2014) <https://slate.com/technology/2014/09/apophenia-makes-unrelated-things-seem-connected-metaphors-paranormal-beliefs-conspiracies-delusions.html#:~:text=He%20was%20describing%20the%20acute,apophany%20is%0a%20false%20realization>.

[252] Peter Brugger, 'From Haunted Brain to Haunted Science: A Cognitive Neuroscience View of Paranormal and Pseudoscientific Thought,' in James Houran and Rense Lange, *Hauntings and Poltergeists: Multidisciplinary Perspectives* (McFarland, 2001) 195; see https://slate.com/technology/2014/09/apophenia-makes-unrelated-things-seem-connected-metaphors-paranormal-beliefs-conspiracies-delusions.html#:~:text=He%20was%20describing%20the%20acute,apophany%20is%0a%20false%20realization.

[253] Sophie Fyfe, Claire Williams, Oliver J Mason and Graham J Pickup, 'Apophenia, Theory of Mind and Schizotypy: Perceiving Meaning and Intentionality in Randomness' (2008) 44(10) *Cortex* 1316, 1317.

[254] Sandra L. Hubscher, 'Apophenia: Defnition and Analysis', *Digital Bits Skeptic* (online at 4 November 2007) <https://archive.md/20130121151738/http://www.dbskeptic.com/2007/11/04/apophenia-definition-and-analysis/#selection-129.0-129.17>.

[255] Ibid.

[256] See Cathy Lovering, 'All About Apophenia', *PsychCentral* (online at 8 December 2021) <https://psychcentral.com/health/apophenia-overview#types>.

[257] Michael Schermer, 'Patternicity: Finding Meaningful Patterns in Meaningless Noise', *Scientific American* (online at 1 December 2008) <https://www.scientificamerican.com/article/patternicity-finding-meaningful-patterns/>.

[258] Ibid.

[259] Cathy Lovering, 'All About Apophenia', *PsychCentral* (online at 8 December 2021) <https://psychcentral.com/health/apophenia-overview#apophenia-examples>.

apophenia, with the behaviour appearing alongside conditions that feature symptoms of psychosis.[260] Interestingly, apophenia may be linked to the 'heightened sensitivity to identifying meaningful patterns in ambiguous stimuli observed in conspiracy believers.'[261]

This helps illustrate the difference between apophenia and pareidolia: With pareidolia, the person perceives the pattern in ambiguous stimuli; with apophenia, the person sees a message or significance for them personally in that pattern. If I see a cloud that I think looks like a love heart, I am experiencing pareidolia. It may make me smile, but nothing more. However, if I see a love heart cloud and think that this is a message from the universe that I should buy my wife roses, I am experiencing apophenia.[262] While this example may be benign, apophenia can quickly become malignant for the gambler chasing the win or the conspiracist convinced of a violent plot against them.

In contrast, pareidolia is now recognised as a normal human tendency[263] that is widely experienced. [264] It has been speculated that the pareidolic tendency to see patterns was an evolutionary behavioural adaptation: as Coolidge and Coolidge note, the 'consequences of a false positive, especially when it involves the recognition of a predator, are not as severe as the consequences of a false negative'.[265] Famously, Carl Sagan, in his 1995 book *The Demon-Haunted World*, argued for this evolutionary benefit:

> As soon as the infant can see, it recognizes faces, and we now know that this skill is hardwired in our brains. Those infants who a million years ago were unable to recognize a face smiled back less, were less likely to win the hearts of their parents, and less likely to prosper … As an inadvertent side effect, the pattern-recognition machinery in our brains is so efficient in

---

[260] Scott Blain et al, 'Apophenia as the Disposition to False Positives: A Unifying Framework for Openness and Psychoticism' (2020) 129(3) *Journal of Abnormal Psychology* 279.

[261] Abdolvahed Narmashiri et al, 'Conspiracy beliefs are associated with a reduction in frontal beta power and biases in categorizing ambiguous stimuli' (2023) 9(10) *Heliyon* 19; Katy Waldman, 'It's All Connected', *Slate* (online at 16 September 2014) <https://slate.com/technology/2014/09/apophenia-makes-unrelated-things-seem-connected-metaphors-paranormal-beliefs-conspiracies-delusions.html#:~:text=He%20was%20describing%20the%20acute,apophany%20is%20a%20false%20realization>.

[262] The difference has been explained in the following way: 'if someone sees something that looks to them like a UFO in the sky, it is pareidolia, but if that same person believes the UFO has chosen them as a subject for experimentation, or maybe as a means to communicate with the human race, then that is apophenia combined with pareidolia. … If someone sees an image of Jesus Christ on their toast, that is pareidolia, but if they then go on to believe that it is God's way of giving them a message, then that is apophenia again.': Ian, 'Pareidolia and Apophenia Explained', *Owlcation* (online at 15 December 2023) <https://owlcation.com/stem/Pareidolia-Explained>. Note this distinction is not always maintained. For example, Zusne and Jones appear to conflate pareidolia and apophenia when they argue: 'Pareidolia are the basis of all those divination practices that involve the visual inspection of patterns formed by random processes, be it tea leaves, smoke, patterns formed by randomly falling objects, or the shapes assumed by molten wax or metal poured into water. Because the processes underlying the formation of these patterns are unpredictable, it is assumed that they are therefore malleable by paranormal influences. When meaningful configurations do emerge, they are assumed to carry a message from a supernatural being or entity or to have been produced by unknown forces': Leonard Zusne and Warren H. Jones, *Anomalistic Psychology: A Study of Magical Thinking* (Psychology Press, 1989) 77.

[263] Rebecca J. Rosen, 'Pareidolia: A bizarre Bug of the Human Mind Emerges in Computers', *The Atlantic* (online at 7 August 2012) <https://www.theatlantic.com/technology/archive/2012/08/pareidolia-a-bizarre-bug-of-the-human-mind-emerges-in-computers/260760/>.

[264] Colin J. Palmer and Colin W. G. Clifford, 'Face Pareidolia Recruits Mechanisms for Detecting Human Social Attention' (2020) 31(8) *Psychological Science* 1001.

[265] Frederick L. Coolidge and Melissa L. Coolidge, 'Why People Se Faces When There Are None: Pareidolia', *Psychology Today* (online at 9 August 2016) <https://www.psychologytoday.com/au/blog/how-to-think-like-a-neandertal/201608/why-people-see-faces-when-there-are-none-pareidolia>.

extracting a face from a clutter of other detail that we sometimes see faces where there are none.[266]

As a consequence of this evolutionary imperative, the human brain has a hard-wired tendency to not only recognise patterns, but to ascribe personality and social meaning: as Palmer notes, 'the windows of a house might feel like two eyes watching you, and a capsicum might have a happy look on its face.'[267] Specific mental content, including emotional states,[268] is ascribed to random patterns seen on inanimate objects.[269] For many researchers, these phenomena underlie human tendencies to believe in the supernatural – to see ghosts and demons.[270]

Pareidolia is recognised to occur in several forms. The most commonly experienced form is 'face pareidolia', which involves the perception of faces in random visual stimuli.[271] Face pareidolia is a subset of the broader form of 'visual pareidolia' – think the animal in the cloud. Another form is 'auditory pareidolia', which occurs when random sounds are perceived as meaningful communication. One example of this is the electronic voice phenomena (EVP), whereby 'electronic recording devices are alleged to have captured audio of human-like voices', which are then 'interpreted as messages from paranormal or discarnate entities.'[272] Another example is the widespread concern in the early 1980s around 'backwards masking' or 'backmasking': that 'rock records allegedly contained satanic messages that could be heard when the record was played backwards.'[273]

In the following section, I argue that we should see the way humans process the phenomena of both pseudolaw and pseudo-language as a form of 'conceptual pareidolia'. Rather than having an illusion created by our visual or auditory processing systems, these phenomena appear to short-circuit the way our brain processes complex social interactions such as language and normative systems.

### (b) 'Conceptual Pareidolia': Pseudolaw and Pseudo-language

In the start of this Part, I argued that both the pseudo-language of LLMs and contemporary pseudolaw possess two common characteristic: firstly, they both mimic the form of the underlying counterpart 'primary' phenomenon (law and language), but not the necessary substance that underpin them, and secondly users mistake the mimicking 'secondary' phenomenon for the

---

primary counterpart (pseudolaw as law; pseudo-language as language). This tendency can, I believe, be helpfully understood by reference to the idea of 'conceptual pareidolia,' the erroneous perception of meaningful patterns in complex social interactions.[274]

The proposition is that in both cases, users perceive a meaningful pattern arising from relevantly meaningless stimuli in a manner that reflects the core definition of pareidolia. The ritualistic behaviours and arguments of pseudolaw are substantively nonsense in the legal context, but adherents believe that they are performing legally significant actions. LLMs mimic the syntax of written language but operate in a closed system, with no intentionality nor externality of meaning, lacking any core semantics or capacity to transfer meaning. In both cases, the mimicking phenomenon bears a surface veneer of similarity with the underlying counterpart phenomenon, but not the functional substance, yet the user mistakes the one for the other. The form is mistaken for the substance, the stimuli perceived as the noumenon. This is pareidolia – yet the deceived cognitive systems are not visual or auditory but conceptual. It is the same Type I error but in a different context.

This framing may help to understand the deep commitment of users to the substantive belief that the perceived pattern (pseudolaw as law; pseudo-language as language) is the thing itself. Pareidolia is hardwired into the human brain, a deeply embedded means of processing partial information quickly and efficiently. As with the perceived tiger's face in the long grass, the 'cost' of treating the false positive as a true positive until proven otherwise is often less than the cost of the false negative. If this pathway *is* activated, then users are likely to believe that the perceived phenomenon is the underlying counterpart phenomenon until the illusion is shattered for them.

This may be the second implication of this framing. In both cases, the underlying counterpart phenomena are inherently complex despite appearances of simplicity. While we all constantly *use* language, most of us are unable to describe how the system operates and how syntax, semantics, semiotics and pragmatics interact to convey and communicate meaning. We are users of language, not linguists. Something that shares the form of language will probably be treated as language until proven otherwise – and given the irreducible complexity and inherent inexplicability of LLMs, for many, no explanation will be capable of displacing that illusion. Similarly, while we are all familiar with the role that law plays in society and have a basic functional understanding of the major norms, most citizens are functionally legally illiterate.[275] Law is notoriously inaccessible for the public, and even legal theorists struggle to define the core nature and distinguishing features of law. In this context, it is astonishingly difficult to explain why *this* recitation of formal language, obscure source and archaic ritual is meaningful while *that* recitation of formal language, obscure source and archaic ritual is nonsense. Effectively piercing the illusion in a manner that destroys it depends upon being able to understand the difference between mimicry and core social phenomenon. For both the pseudo-language of LLMs and pseudolaw, the nature of both the primary and secondary phenomenon makes this particularly hard.

Understanding the use of LLMs and pseudolaw as forms of conceptual pareidolia helps to illustrate that: (1) in both cases an illusion of meaning is created from nebulous stimuli; (2) the human mind is hard wired to identify such patterns; (3) the ability to shatter the perceived illusion requires a shift of perception that may depend upon an internalised understanding of the underlying primary phenomenon. This framing can help us understand not only the successes of the secondary phenomenon but also their apparent resilience.

---

[274] It is important to note that while I describe the practices above as a form of 'conceptual pareidolia', it is sufficient for these purposes that this be as an analogy rather than substantive cognitive phenomenon. Whether or not similar neural processing pathways are deployed, and deceived, in processing pseudo-language as in the case of language itself is an issue that may warrant further analysis but which is far beyond the scope of this paper. In understanding the apparent success and durability of these secondary phenomena, the analogy is useful in revealing the core illusion of meaning

[275] See below Part V(2)

This framing also helps to shift the focus from the *generation* of the illusion, the *perception* of it. As I outlined in Part II, the critical point in the success of LLMs in not whether these technologies are stochastic parrots or not, but rather whether they are perceived as conveying meaning. Bender herself recognised that perceptions of language are affected by our natural linguistic competence, observing:

> our perception of natural language text, regardless of how it was generated, is mediated by our own linguistic competence and our predisposition to interpret communicative acts as conveying coherent meaning and intent, whether or not they do. [276]

Yet, once we understand the secondary phenomenon of pseudo-language through the framing of pareidolia, we see that this mediation by our inherent linguistic competence may create a processing error – lacking the relevant context, we mistake the secondary mimic phenomenon for the primary phenomenon of language. We perceive the illusion as being a communicative act of 'coherent meaning and intent'.

Our focus should be on the audience member perceiving the illusion, not on the illusionist. The framing of conceptual pareidolia helps highlight this.

## 2. Confidence as a Proxy for Competence

Secondly, part of the efficacy of both phenomena in creating illusions that mimic the underlying primary phenomenon is that the relevant outputs are presented to the audience with a confidence that belies their ethereal lack of substance. In both cases, there appears to be an activation of a heuristic in the way humans process information when it is presented with confidence: the 'confidence heuristic'. As I describe below, when humans read/hear confident-sounding words, lacking qualifiers of doubt, we tend to assume a degree of expertise by the writer: *confidence* is treated as a proxy for *competence*.

This tendency is evident in the way users interact with both LLMs and pseudolaw. When LLMs produce content, they do so in a manner that presents that output normal qualifiers of doubt. The output is *confident, so we assume the model is competent*. As Hicks et al note, LLMs are 'simply not designed to accurately represent the way the world', but – critically – are designed 'to *give the impression* that this is what they're doing'.[277] The output is presented as if it is substantively correct, even when it lacks the foundational conceptual design to ensure that this is the case. Allardice describes this as the 'Confident Wrongness' of ChatGPT,[278] and the mode of presentation is leading to significant concerns about the overreliance on the technology.[279] Even before the rise of LLMs, the research demonstrated that humans over-trust the AI:[280] people 'supported by AI-powered decision support tools frequently over-rely on the AI.'[281] In the context of AI-advised human decision-making, humans tend to rely on AI-generated (but incorrect) predictions 'even when they would have made a better decision on their own.'[282] Essentially, people appear to have developed a 'heuristics about the competence of the AI partner overall.'[283] Generative AI appears to have turbocharged this bias, as the confident form of the output is affecting the ascribed reliability of the output.

Similarly, users of pseudolaw are assertive and convinced of their correctness when in court, lacking the normal hesitations of LIPs. They *present as confident* because *they believe they are competent.* In our recent empirical study of pseudolaw,[284] several judicial officers observed this confidence of users:

> They will come a lot of them and have this confidence what they are preaching, their argument, it's like they've studied this script and they come to court and then they just reel it off. They have this confidence[285]

The adherents are not only self-deceived about their own competence, but they are also deceived by the confidence with which gurus and other pseudolaw adherents present knowledge. Pseudolaw mimics the forms of law, and does so confidently. Users believe, therefore, that they are performing law.

### *(a) Bullshit and Pseudo-Language*

It has been suggested that what is occurring here is a form of structural 'bullshit'. For example, Hicks, Humphries and Slater, have memorably argued that ChatGPT is 'a bullshit machine,'[286] that is 'designed to produce text that *looks* truth-apt without any actual concern for truth.'[287] The authors here are drawing upon the influential concept of 'bullshit' as developed by Harry Frankfurt in his 2005 book *On Bullshit*. In that work, Frankfurt describes 'bullshit' in the following way

> The fact about himself that the bullshitter hides … is that the truth-values of his statements are of no central interest to him; what we are not to understand is that his intention is neither to report the truth nor to conceal it. This does not mean that his speech is anarchically impulsive, but that the motive guiding and controlling it is unconcerned with how the things about which he speaks truly are[288]

In this way, bullshit is distinguished from lying. While the liar and the bullshitter both 'represent themselves falsely as endeavouring to communicate the truth',[289] only the liar 'wants us to believe

---

something he supposes to be false'.[290] The bullshitter has no concern at all with the truth: the defining feature of bullshit is "*a lack of concern* with truth, or an *indifference to how things really are*".[291]

Hicks, Humphries and Slater build upon this framework to distinguish between 'Hard Bullshit' and 'Soft Bullshit' depending upon whether the utterer has an mislead the hearer.[292] They go on to propose that, at a minimum, the output of LLMs are soft bullshit 'speech or text produced without concern for its truth … produced without any intent to mislead the audience about the utterer's attitude towards truth.'[293] They argue that this approach of conceiving ChatGPT as a bullshit machine is beneficial, as it places the focus on the (un)reliability of the output:

> This is why we favour characterising. This terminology avoids the implications that perceiving or remembering is going on in the workings of the LLM. We can also describe it as bullshitting whenever it produces outputs. Like the human bullshitter, some of the outputs will likely be true, while others not. And as with the human bullshitter, we should be wary of relying upon any of these out[294]

Given the way I have described LLMs above as producing illusions of meaning, it should not be surprising that this framing is attractive. The focus on the bullshit quality of the output helps to highlight the inherent unreliability of the outputs. Yet, as with Bender's work, this characterisation focusses attention on the generation of output and not its reception/perception. LLMs may indeed produce bullshit. But what is particularly interesting about them is that humans do not perceive them as producing bullshit, but rather seem predisposed to accept the accuracy and truth of the outputs. It is the fact that LLMs mimic the forms of language and behaviour of 'truthiness' that makes them effective, not merely that they are agnostic/indifferent to the truth. When humans read confident-sounding words, lacking qualifiers of doubt, we assume a degree of expertise by the writer. LLMs apes this form. This goes beyond bullshit.

How then should this be understood? Rafferty has suggested that what is occurring is a form of "Epistemic Impersonation",[295] where the form of epistemic authority is being utilised without the substance. There is something attractive about this concept. Yet impersonation denotes agency by the impersonator, while LLMs are not agents and lack all intentionality. With LLMs, the relevant inference occurs on the part of the reader.

A better way of understanding the success of both pseudolaw and LLMs is, instead, to focus on the way in which humans process information and how the confidence with which information is presented acts as a proxy for the reliability/persuasiveness of that information. That is, the form in which both pseudolaw and pseudo-language is presented may be triggering the 'confidence heuristic' in the human brain.

### *(b) The Confidence Heuristic*

---

The idea of a confidence heuristic was first developed by Thomas and McFadyen in a 1995 paper *The Confidence Heuristic: A Game-Theoretic Analysis*.[296] In that paper, they described the novel heuristic 'by which individuals try to assess the reliability of information'[297] arguing that:

> 'Individuals may base their decisions about the reliability of information upon the confidence with which it is expressed: *they judge that someone is more likely to hold reliable information if he or she expresses that information confidently* rather than tentatively.[298]

As the authors note, like all heuristics, 'the confidence heuristic is a short-cut to deciding conditions of uncertainty and is prone to error.'[299] Nevertheless, it is now well recognised that the confidence with which information is presented affects its persuasiveness. Interestingly, research suggests that 'verbal rather than nonverbal communication drives the heuristic'.[300]

As I discussed earlier, both phenomena in this paper are distinguished by a particularly confident presentation of their content to users. The confidence heuristic can help explain why users are so ready to believe that the output of both phenomena is substantively meaningful. The confident *form* tricks our brains into accepting the accuracy of the information. In Part II and III of this article, I show that both phenomena utilise the form but not the substance of the underlying primary phenomenon. Here, we see that the confidence with which that form is adopted that may be significantly responsible for the efficacy of mimicry.

The advantage of identifying the role that the confidence heuristic may be playing here is that, like all heuristics/cognitive biases, the aware we are of the bias, the better able we are to counter it. It is not merely that LLMs are creating bullshit. It is that they create confidently presented bullshit that trigger a cognitive bias in out brain that leaves us more vulnerable to accepting the substantive truth of the proposition. Likewise, when a pseudolaw adherent reads pseudolegal materials and arguments developed by gurus or is presented with a comprehensive script to present in court, the very confidence by which those materials mimic legal forms makes that adherent more likely to believe that what is presented is true and is 'true law'.[301] Pseudolaw may be nonsense, but it is *confident* nonsense.

When we acknowledge that cognitive bias is potentially at play, this should affect the way in which we respond to both the relevant output and the users of such outputs. While courts may already recognise that many pseudolaw adherents do actually believe the arguments they put forward,[302] this framing can help them understand the resilience of that belief and also to explain to adherents where that belief may emanate. Similarly, this framing can help all users of LLMs to be critical in their reception of the information and recognise not only that the output may be unreliable bullshit, but that we humans are particularly bad at identifying this as a result of the confidence of the form.

---

[296] Jonathan P. Thomas and Ruth G. McFadyen, 'The confidence heuristic: A game-theoretic analysis' (1995) 16(1) *Journal of Economic Psychology* 97.

[297] Ibid, 100.

[298] Ibid, 100.

[299] Ibid.

[300] Briony Pulford et al, 'The Persuasive Power of Knowledge: Testing the Confidence Heuristic' (2018) 147(10) *Journal of Experimental Psychology: General* 1431-1444: This may help explain the relative success of both LLMs and pseudolaw given that both heavily utilise the written word.

[301] Joe McIntyre et al, *The Rise of Pseudolaw in South Australia: An Empirical Analysis of the Emergence and Impact of Pseudolaw on South Australia's Courts* (Final Report, University of South Australia, September 2024).

[302] See for example, *Taylor, In the matter of an application for leave to issue or file*, [2023] HCA Trans 63 (18 May 2023)

### 3. Success as a Triumph of Form

Thirdly, this awareness of form over substance should help us identify the ways in which both phenomena can be successful – despite their substantive limitations. Because it is important to recognise the ways in which both phenomena are successful, just not in the ways in which they are often presented as being successful.

#### (a) Failures in their Claimed Domain

But first, it is worth noting that by the metrics by which they are conventionally understood, both LLMs and pseudolaw are not successful.

Pseudolaw presents itself as offering a means of countering legal powerlessness by offering a set of tools that promise to be legally effective. Yet these tools categorically do not work; no pseudolegal argument has ever been successful anywhere in the world.[303] Pseudolaw, understood as a means of effective legal argumentation, is an abject failure. In those rare cases where a pseudolaw litigant wins, they do so on conventional legal grounds.[304] As means of successfully and substantively engaging with the law, pseudolaw is phenomenal unsuccessful.

Similarly, LLMs have been portrayed as fostering a new 'AI Revolution' that is 'transforming industries'.[305] Ethan Mollick, writing soon after the launch of ChatGPT in late 2022, wrote that the technology will 'quickly and accurately generate written content' in a way that allows businesses to 'save time and resources … [and] to focus on other important tasks.[306] Sam Altman, CEO of OpenAI has been a key player in advancing this hype, suggesting for example that:

> AI is going to eliminate a lot of current jobs, and there will be classes of jobs that totally go away. AI is also going to change the way a lot of current jobs function, and it's going to create entirely new jobs.[307]

However, these promises have not come to fruition and appear hyperbolic. OpenAI itself faced losses of over US$5billions in 2023/24,[308] despite having over 200 million weekly users.[309] Large

---

technology companies invested over US$200 billion in data centres for AI[310] only to be embarrassed by Deepseek providing disrupting the industry with cut-price offerings.[311] There is now growing scepticism about the ability of LLMs to offer any meaningful competitive advantage,[312] and there have been no 'killer apps' that have emerged.[313] By the end of 2024, technologist Ed Zitron suggested, provocatively, that perhaps 'Godot Isn't Making it'.[314] This lack of adoption by industry is reflective of the inherent unreliability of the technology. While the accuracy of LLMs is an area of intense academic analysis, some research suggests that 30-40% of outputs may involve hallucinations[315] with factual errors present in 46% of generated texts.[316] While in many instances, LLMs are exceptionally good at producing accurate and compelling outputs, they are also prone to getting issues catastrophically wrong. The technology is unreliably accurate. And because of the black-box nature of the technology, there is no explicability of why a given answer is given, nor means of predicting when a hallucination will occur. The technology is generally pretty good, rarely excellent, and sometimes terribly wrong. In a 2023 study, Kocoń *et al* attempted to assess the accuracy of the technology and concluded:

> 'ChatGPT can solve most of the problems considered quite well. On the other hand, it [consistently] loses to the best models currently available … Its loss is relatively greater for more difficult and pragmatic tasks.... All this makes ChatGPT a master of none of the tasks'[317]

To the extent that LLMs promise accurate and reliable substantive content, they should be seen as failures. Businesses, particularly knowledge industries, do not want a product that 'mostly kind of ok, but sometimes, unpredictably, woefully wrong'. Such unpredictability, particularly when combined with the confidence heuristic discussed above, is not consistent with a high-value product that clients will want to pay for. And it is already clear that people are not willing to pay for access to LLMs.[318] For example, Windows is now forcing unwilling subscribers to pay for a premium bundled AI service[319] as they seek to recoup investments in the technology. While LLMs

---

seem to promise much, it already appears that the initial promises of revolution were overblown. While the mimicry of the form is important, ultimately, it is the substance for which people are willing to pay.

### (b) Success When the Form Matters

However, this does not mean that these phenomena are abject failures. Rather, it focuses our mind to home in on those spaces where they do succeed. While both pseudolaw and pseudo-language are substantive failures, both achieve a measure of success when we shift our focus to the foregrounded form. Both phenomena are highly developed illusions of the mimicked primary phenomenon, and in some cases, the creation of a successful illusion is sufficient to achieve the desired outcome.

With pseudolaw, the mimicry of form does allow a measure of success in two distinct ways. Firstly, the form-focused presentation of *pseudolaw as law* does appear to succeed for the various 'gurus' who seek to monetise access to pseudolegal knowledge. There is a long history of such gurus charging significant fees to adherents to gain affiliation,[320] access to website and repositories,[321] or to attend events.[322] In these instances, the fact that pseudolaw looks sufficiently like actual law is likely to be a critical component in convincing targets to pay. Secondly, there is evidence that pseudolaw performances can achieve a measure of success in forcing other participants in court proceedings to alter their behaviour by challenging courtroom hierarchy and disturbing normal court processes.[323] The behaviours of adherents in that study presented a distorted mimicry of litigious behaviour, and while legally unsuccessful,[324] were effective in establishing a 'communicative expertise' that subverted the schematic order of proceedings.[325] When 'success' is defined in terms other than substantive legal success, then pseudolaw can be capable of achieving beneficial outcomes for adherents. The more closely aligned that success is to the form of the behaviour the greater the chances of success.

This same pattern is evident with LLMs. While these models are inherently unreliable when one focuses on the substantive meaning of outputs, they remain highly effective at producing pseudo-language that mimics the form of written language. As a result, LLMs will succeed in those situations where users *don't actually care about the accuracy of the substantive content*. There are many applications where what is desired is a written output that looks a particular way, but where the accuracy of the content is unimportant. One court (perhaps uncharitably) suggest situations such as writing corporate mission statements, or university policy. More contentiously, LLMs appear to be particularly well suited to producing written student assessments – even in fact

---

dependant disciplines such as law.[326] At first glance, this seems anomalous, as such assessments seem to be substance heavy. However, assessments – and, importantly assessment marking guidelines – largely rely upon form as a proxy for substance and clarity of expression as a proxy for knowledge. It has been suggested that AI is a 'plagiarism machine',[327] and there is much to commend this view. But this should not be taken as proof that LLMs are good, for example, at writing persuasive legal prose, so much as evidence that assessments focus largely on form. In short, if an LLM is producing an output that seems particularly helpful, the user is likely more interested in form than substance. And this is not a form of success that is ever likely to generate significant income.

## 4. Magical Thinking

Finally, the focus on form helps to highlight one of the possible reasons for the 'success' of both phenomena (in terms of convincing users that they are substantively meaningful): both make use of 'magical thinking', with users wanting to believe in the magic on offer. Essentially, because the form is so good, the user thinks that output/performance is substantively meaningful *because they want it to be meaningful*.

### (a) The Insurmountable Problem and the Magical Solution

Both pseudolaw and LLMs promise an accessible and powerful solution to a similar problem:

1. Users are presented with an almost incomprehensibly large information-rich environment;
2. The development of substantively meaningful and effectively tailored individual response in such a context depends upon the ability to access, comprehend and synthesise that information, which in turn depends upon a range of literacies (technological, legal, domain-specific etc);
3. For that analysis to be effective, it must not only be tailored, but it must be presented in a particular form, which again requires relatively advanced skills;
4. For many people, not only do they not possess these skills, but they do not have the resources to obtain expert assistance;
5. Taken together, these factors create a sense of powerlessness and alienation, which diminish the perceived agency of the user.

For pseudolaw, this problem is relatively explicit. Law is notoriously inaccessible, and regulatory restrictions generally limit publicly available legal information to largely generic summaries. Yet individuals are regularly faced with serious legal problems, only to find themselves not only unable to afford a lawyer, but they also lack any effective legal literacy.[328] Faced with this powerlessness and alienation, pseudolaw and its promise of empowerment is instantly appealing.[329] Pseudolaw offers the promise of the democratisation of law.

---

[326] See for example, Jonathan Choi, Kristin Hickman, Amy Monahan and Daniel Schwarcz, 'ChatGPT Goes to Law School,' (2022) 71(3) *Journal of Legal Education* 387; Miriam Sullivan, Andrew Kelly, Paul McLaughlan, 'ChatGPT in Higher Education: Considerations for Academic Integrity and Student Learning' (2023) 6(1) *Journal of Applied Learning and Teaching* 31–40; Damian Curran, Inbar Levy, Meladel Mistica & Eduard Hovy, 'Persuasive Legal Writing Using Large Language Models' (2024) 34(1) *Legal Education Review* 184; Amanda Head and Sonya Willis, 'Assessing Law Students in a GenAI World to Create Knowledgeable Future Lawyers' (2024) 31(3) *International Journal of the Legal Profession* 293

[327] Homto Dokpesi, 'Is Generative AI A Plagiarism Machine?', *Medium* (online at 6 October 2023) <https://medium.com/illumination/is-generative-ai-a-plagiarism-machine-5f1e5846a049>.

[328] See generally, Joe McIntyre, The Ur-Controversy of Civil Courts (2022) in Marg Camilleri and Alistair Harkness (eds) *Australian Courts: Controversies, Challenges and Change* (Springer, 1st ed, 2022) 345.

For LLMs, this problem is more implicit. While the advent of the internet has made vast repositories of human knowledge widely available, much of that knowledge remains behind barriers, both financial (eg paywalls) and conceptual (domain expertise and literacy necessary to understand and process). The underlying data may be available, but it is commonly not usable. Usability demands a capacity to access, understand, synthesise and functionally effectively deploy that information. LLMs seem to promise the user the ability to access this vast set of knowledge and to quickly and accessibly obtain tailored and specific substantive information that is presented in a persuasive and usable form. LLMs offer the promise of the democratisation of knowledge.

Faced with this promise, and an output that looks like it delivers on that promise, users are understandably willing to believe that the promise is being met. Yet, ultimately, what is actually occurring is an appeal to magic: faced with an impossible task, an easy ritual bypasses the challenge and effortlessly delivers the desired result. I rub the lamp, and the genie grants my wish – I have a powerful legal argument, I have a beautifully written summary of a book, a report, a complex geo-political event. The user is instantly granted access to knowledge and skills that they have not had to learn through traditional efforts. Magic is, of course, innately appealing.

### (b) Magical Thinking in Pseudolaw and Pseudo-Language

Strikingly, the language of 'magic' is extensively used with respect to both phenomena. With pseudolaw this framing has been prominent for at least a decade. In his influential judgment in *Mead v Mead,* Rooke J observed that pseudolaw is a 'drama that is more akin to a *magic spell ritual* than an actual legal proceeding.'[330] More recently, scholars such as Pitcavage have explicitly that '[m]agical thinking is at the heart of most sovereign citizen pseudolaw'.[331] Pitcavage explains this aspect of pseudolaw in the following way:

> 'Magical Thinking' occurs when people believe that specific words, rituals or actions can affect events in the physical world. Just as a shaman might perform a ritual to heal someone afflicted with a disease, so too do sovereign citizens invoke court cases, statutes and regulations, and legal phrases to free themselves from laws, financial obligations or other unwanted manifestations of government or authority[332]

This aspect of pseudolaw even has its own nomenclature: 'lexomancy'. Griffin has coined the term to refer to the way in which pseudolaw utilises 'legitimate legal signs … in an effort to

---

talismanically imbue texts with real authority'.[333] Griffin has subsequently provided a useful expansion of the magical explanation of pseudolegal in the following terms:

> Sovereign Citizens are generally individuals who do not have a strong understanding of the structure of the legal system or the way it operates. In their encounters with its representatives (i.e. lawyers, police, judges, etc.) Sovereign Citizens see those representatives perform legal rituals that they do not comprehend and that have dramatic real word effects (e.g they receive a traffic ticket, are arrested, or have their homes foreclosed upon). Whet consciously or not, because of their lack of understanding (or perhaps because of a willful misunderstanding) of how the legal system functions, Sovereign Citizens conclude that those legal rituals tap into some element of the "supernatural" and decide to attempt to claim that power for themselves to turn it against their oppressors. *To bolster their efforts in what they perceive to be a form of magical combat, Sovereign Citizens do not just coop the form of existing legal rituals; instead, they make efforts to enhance what they believe to be the most magically salient features.*[334]

It is critical to note that all these expositions of the role of magical thinking in pseudolaw forefront the role of the form of pseudolaw. Pseudolaw may be 'obvious nonsense'[335] to lawyers, but that is almost beside the point. Adherents employ ritual and modes of behaviour in a manner they think is legally meaningful, and in doing so, they think they can become legally powerful. This is the promise of lexomancy. Something impossible becomes achievable: 'Suddenly, law is mine to control, I can get what I want right now!'

Similarly, the idea of magic thinking has been a familiar trend in the analysis of LLMs. Leaver and Srdarov, in an article explicitly addressing the idea of technological magic observe that for many people, 'the combination of AI technologies and media hype means generative AIs are *basically magical* insomuch as their workings seem impenetrable, and their existence could ostensibly change the world'.[336] Certainly, how the technology has been discussed in public discourse reflects this understanding. One journalist opined that 'ChatGPT is (Sometimes) Indistinguishable from Magic',[337] while another, writing in early 2023, that the technology results in a 'kind of magic' as 'machines that have ingested an internet's worth of data, weighed up the relationships between things, and are able to generate content that appears to be new and original.'[338] As Leaver and Srdarov note, the companies that offer LLMs tend to lean into this framing, positioning 'their products as seemingly magical' to make them 'even more appealing to potential customers and investors'.[339] Perhaps nowhere is this magic-themed hype more explicit than in the marketing of the *AI Avalanche* subscription service 'ChatGPT Black Magic', which claims to train users in better utilising the technology and uses the catchphrase 'You'll learn how to achieve things so incredible, they <u>will seem like witchcraft</u>.'[340] LLMs seem to offer a way for users to gain meaningful and useful access to the entirety of human knowledge. What could be more magical and appealing? That it is an illusory promise is beside the point. The form is convincing, and the appeal is irresistible. We should not be surprised that users wish to believe in the magical promise, and are often committed to that belief.

### (c) Responding to Magic

Author C Clark famously observed that '[a]ny sufficiently advanced technology is indistinguishable from magic'.[341] Both pseudolaw and LLMs thrive in part because they appear to offer a form of technology (legal/computational) that is advanced and relevantly incomprehensible, and yet is substantively powerful. For the users, this is effectively magic. And the promise of magic is beguiling. It offers a means to bypass insurmountable barriers, a way for the weak to become powerful. Understood in this way, it is understandable that users appear committed to the mistaken belief that the success of the relevant form is indicative of the creation of reliable and meaningful substantive success.

Yet, by better understanding the distinction between successful form and successful substance, we are better able to identify that what is occurring is simply magical thinking. In turn, this helps us to understand the nature of the challenge posed by the rise of these phenomena and to tailor responses accordingly. In their analysis of generative AI, Leaver and Srdarov observed that:

> Escaping the hype and hypocrisy deployed by AI companies is vital for repositioning generative AI not as magical, not as a saviour, and not as a destroyer, but rather as a new technology that needs to be critically and ethically understood.[342]

LLMs succeed in part because of the alienation of the population from much of human knowledge and the skills necessary to access it. These technologies are incredibly good at mimicking the *form* of substantively meaningful content, and beguile people into believing that they are therefore, delivering such content. Because of that alienation, people want to believe in that magical promise. If we are to critically and ethically engage with this phenomenon, it is not only necessary to understand what is and is not being produced, but also the ways that humans (in all their glorious irrationality) engage with that promise.

---

In the same way, courts and legal institutions must understand that pseudolaw is largely perceived as magic, even if it is nothing but an ethereal illusion. It is not enough to point out to adherents that this is legally ineffective. Rather, the institutions of law must be willing to recognise that the magical appeal of pseudolaw flows from policy choices made in these institutions, choices to tolerate alienation and poor legal literacy, and failure to act on maintaining meaningful access to justice. The magical thinking of pseudolaw is so appealing precisely because the alternative powerlessness is so overwhelming and insurmountable.

## PART V: CONCLUSIONS

In this article, I have sought to develop two key observations about the emerging social phenomena of pseudolaw and LLMs. Firstly, I argue that both phenomena elevate form and appearance over substance and content. Pseudolaw produces outputs that *look like* legal reasoning and legal language and utilise rituals and behaviours that appear to be legally significant and meaningful. Yet this is illusory, with pseudolaw fundamentally disconnected from the content of law, and entirely legally ineffective. Similarly, LLMs produce outputs that *look like* they are meaningfully and intentionally reasoned summations of externally validated information – that its output is both syntactically and semantically meaningful. Yet, again, this is illusory, with a combination of brilliant engineering, phenomenally large datasets and statistical calculations of hundreds of billions of permutations producing a highly developed simulacrum of language, effectively a pseudo-language. This first observation is essentially technical in nature – looking at the quality of the jurisprudential nature of the pseudolaw as a matter of 'legal science', or at the quality of LLMs as a matter of computer science.

Secondly, I argue that users of these technologies routinely mistake the form for the substance: pseudolaw adherents genuinely believe they are making legally meaningful arguments, that they are 'doing law; users of LLMs routinely overestimate the nature of the pseudo-language output, mistaking it for meaningful language and relying upon it for its substantive content. In both cases, the users are mistaking the illusion of meaning for the object itself. This is not an issue of technical science but of behavioural science – it is a question of how humans understand and respond to illusory perceptions.

In the second half of this article, I explore the implications of this conception of the two phenomena and explore how this recognition of triumphant form over substance helps us understand the appeal and limitations of both phenomena. I argue that the apparent success of both phenomena may be a consequence of deeply embedded human tendencies regarding the identification of patterns from nebulous inputs. I argue that in both cases, there appears to be the operation of 'conceptual pareidolia', the erroneous perception of meaningful linguistic/legal patterns from nebulous inputs. Secondly, I argue that both appear to rely upon the 'confidence heuristic', whereby the human cognitive bias for treating confidence as a proxy for competence is activated by the way in which these illusions of meaning are presented to reinforce the belief that the outputs are meaningful and significant. Thirdly, I argue that this elevation of form over substance helps to highlight the ways in which both phenomena can succeed, namely when the concern is with the form of the output and not its content. Pseudolaw is effective for the gurus and charlatans who sell access to the apparently legally significant arguments and for users who wish to disrupt the rituals of legal proceedings. LLMs remain incredibly effective for the generation of form-rich/content-poor text, where the focus is on how language is used rather than its communicative intent. Finally, I argue that both phenomena draw heavily upon the magical thinking of users and the desire for the promise of the technology to be real. Both offer the potential to democratise access to arcane and extensive knowledge that is otherwise inaccessible. We should not be surprised that users wish to avail themselves of these apparent magical tools. Yet it is important to recognise that it is magical thinking

## 1. Law as an Analytic Lens

There are good reasons why I draw together these two particular phenomena in the context of law. The nature of law as a technical and often esoteric enterprise, that is, language and data rich and which describes itself in logical terms, seems to be an ideal site for the application of LLMs. At the same time, the broken mirror that pseudolaw holds up to law invites reflection on the nature of law itself. Moreover, law offers a particularly rich site to understand the nature of both these phenomena, not least as there is a growing body of evidence that litigants are using LLMs to produce legal documents that increasingly display pseudolegal qualities. We are seeing lawyers and laity alike rely upon LLMs to undertake legal research and produce documents for court proceedings – often with very little appreciation of the limits of that technology. This highlights the overlap between that domain and pseudolaw: both produce outputs that *look* like law, and to the untrained eye may be mistaken for law, but both are capable of generating only (in the technical sense) bullshit.

Both phenomena fail in the context of law because while legal form, ritual and behaviour are important parts of law, they are not law. Law is *system of reasoning*, a *system of sources,* and a political *system of social governance*. While its logical and legalistic form is important, it is important because of the way in which it enables users to use and manipulate legal norms in specific, explainable and tailored ways to concrete situations. Law may use formal-sounding rules, but it is never statistically predictable because it responds to the infinitely variable nature of human life. This renders the scripts of pseudolaw and the probabilistic modelling of LLMs apostate. Ultimately, law is a discursive enterprise that depends upon one's capacity to reason, persuade and explain. It is dynamic and responsive. It does not give simple answers. It is the antithesis of the black box.

Both LLMs and pseudolaw miss this underlying intellectual infrastructure of law. They produce and output that looks 'lawish', but it is disconnected from law *other than* in the way meaning is ascribed to them by the user. They both encourage users to mistake hallucinations for truth. Both create pareidolic illusions of meaning.

## 2. Responding to Illusions of Meaning

Yet the lens of law is useful for more than just highlighting the nature of the problem. Law can also help highlight the ways in which effective responses to the problem can be developed.

In the legal context, we see that the rise of both LLMs and pseudolaw has arisen in part as a result of the effectively complete alienation of the population from meaningful engagement with the law and its processes – they are the logical, if undesirable, consequences of the lack of access to justice. For a population without understanding of the law, and without meaningful ways of engaging legal expertise, both phenomena appear to offer the promise of *democratising law*. Yet in both cases, the output tricks the citizen into thinking that the product is delivering meaningful engagement with law, while delivering only meaningless drivel.

In the legal context, there is growing recognition that one major part of responding to this issue is to better educate the population about the nature and uses of law; that is, to enhance the legal literacy of the population. The growing literature on the desirability of 'legal literacy' explores the idea that it is necessary to master legal discourse at a level to conduct 'a meaningful and active life in a world saturated with a legal culture'.[343] It involves developing a sufficient set of legal knowledge and skills to be aware of the role of law in everyday life,[344] as well as some capacity to engage with legal issues that arise.[345] It can be defined as:

> the background information stored in one's mind that enables them to take up a legal document, read an article in the newspaper on a law-related event, with an adequate level of comprehension, getting the point, grasping the implications, relating what they read to the unstated context which alone gives meaning to what they read.[346]

The absence of legal literacy has the opposite effect. It retards legal and political engagement, embeds social and economic disadvantage and undermines access to justice. And the more users attempt to engage with law without addressing this underlying lack of legal literacy, the more entrenched this alienation and frustration becomes, and the harder it is to displace the mistaken conceptions of law. Leading Australian constitutional law scholar Emeritus Professor Anne Twomey has recently acknowledged this point. In a submission to a parliamentary inquiry into civics education, engagement and participation in Australia, Twomey notes that she is often contacted by people alleging 'all kinds of constitutional conspiracies and legal "errors" …which magically cause all law to be invalid in Australia and all courts to have no authority'.[347] She explains further:

> The problem is that by the time I try to explain the misconceptions or falsities that are at the root of their arguments, the people making them are so far down the rabbit-hole and so committed to this fantasy world, that they cannot be brought back to reality. The only way that this can be headed off is for Australians, when they are young, to be given a sound understanding of the basics of the system of governance and law, so that they can easily recognise and dismiss pseudo-legal nonsense when they see it. Essentially, we need to be inoculating people by giving them knowledge and the skills to engage in logical reasoning, so they can make a rational assessment of the vast array of material that they are now exposed to on the internet, and discern what is authoritative and sensible as opposed to what is false and manipulative and derived from dubious sources.[348]

Twomey is correct. One way of 'inoculating people' is to ensure that individuals have the knowledge and skills to deal with materials presented to them through pseudolegal networks (and potentially through LLMs).

Because it is impossible to pierce the pareidolic illusion unless we are capable of explaining to people, in ways that is meaningful to them, why their perception is simply showing them an

---

illusion. Without them possessing a working degree of legal literacy, it is literally impossible to explain to adherents why pseudolaw is not law. Such literacy is not some desirable 'extra' that would be nice to have in society. It is the only way we can inoculate the population against the dangers – and resultant social and individual harms – that come with pseudolaw.

And this is as equally true of LLMs. Such models make exceptional use of enormous datasets to produce outcomes that appear astonishingly good. They promise so much and trick us into believing that they deliver on that promise. Yet these models are not producing language (which has an inherent appeal to an externality of meaning, and a necessary intentionality in application) but rather pseudo-language (which merely mimics meaning and intention). But magical thinking and cognitive biases deeply imbedded in the human mind make us want to believe in the illusions they create. As with pseudolaw, it is impossible to pierce the illusion unless one obtains a sufficient degree of technological literacy. It is only once the user possess with such resultant knowledge and skill that it becomes possible to meaningfully dismantle the illusion – to effectively see that there is no tiger in that grass, but a beguiling play of shadows on the straw. The dangers of failing to address this issue may be profound – it is will not be a few foolish lawyers citing made up cases, but Government making profound social choices on the basis of hallucinations.[349] It is ultimately a social necessity that work be undertaken to enhance technological literacy, and specifically literacy about the nature of LLMs, to provide the population with the tools necessary to reveal the illusion, to strip back the hype.

Pseudolaw and LLMs both offer something that is deeply appealing. They promise users that they too can have access to arcane knowledge, they too can be wizards of this new domain without the arduous work of the traditional labourer. Both create illusions of meaning that seem to deliver on that promise. But both are hollow, successful only when form is allowed to triumph over substance. And yet so effective is the illusion, so well targeted to behaviour tendencies in the human mind, that they appear almost impossible to displace. Ultimately, these illusions can only be pierce when the users possess a sufficient legal/technological literacy to enable them to see the illusion as an illusion. There is no short cut to this. It is, though, a challenge we must embark upon.

---

[349] See, eg, Rowena Marshall, 'AI Should Replace Some Work of Civil Servants, Starmer to Announce' *The Guardian* (online, 13 March 2025) <https://www.theguardian.com/technology/2025/mar/12/ai-should-replace-some-work-of-civil-servants-under-new-rules-keir-starmer-to-announce>.